\documentclass[reprint,showkeys,amsmath,amssymb,aip]{revtex4-2}
\usepackage[T1]{fontenc}
\usepackage[utf8]{inputenc}
\usepackage{placeins,graphicx,dcolumn,bm,xcolor,soul}
\usepackage[hidelinks]{hyperref}
\usepackage[caption=false]{subfig}

\begin{document}
\preprint{APS/123-QED}

\title{Fractal and Wada escape basins in the chaotic particle drift motion in tokamaks with electrostatic fluctuations}

\author{L. C. de Souza} 
\thanks{Corresponding author: leonardo@fisica.ufpr.br}
\affiliation{Universidade Federal do Paraná, Departamento de Física, Curitiba, PR 81531-990, Brazil.} 
\author{A. C. Mathias} 
\affiliation{Universidade Federal do Paraná, Departamento de Física, Curitiba, PR 81531-990, Brazil.} 
\author{I. L. Caldas}
\affiliation{Universidade de São Paulo, Instituto de Física, Departamento de Física Aplicada, São Paulo, SP 05315-970, Brazil.} 
\author{Y. Elskens}
\affiliation{Aix-Marseille Université, PIIM, UMR 7345 CNRS case 322 campus Saint-Jérôme, F-13397 Marseille, France.}
\author{R. L. Viana}
\thanks{Partially supported by FAPESP}
\affiliation{Universidade Federal do Paraná, Departamento de Física, Curitiba, PR 81531-990, Brazil.} \affiliation{Universidade de São Paulo, Instituto de Física, Departamento de Física Aplicada, São Paulo, SP 05315-970, Brazil.} 

\date{\today}

\begin{abstract}
The ${\bf E}\times{\bf B}$ drift motion of particles in tokamaks provides valuable information on the turbulence-driven anomalous transport. One of the characteristic features of the drift motion dynamics is the presence of chaotic orbits for which the guiding center can experience large-scale drifts. If one or more exits are placed so that they intercept chaotic orbits, the corresponding escape basins structure is complicated and, indeed, exhibits fractal structures. We investigate those structures through a number of numerical diagnostics, tailored to quantify the final-state uncertainty related to the fractal escape basins. We estimate the escape basin boundary dimension through the uncertainty exponent method, and quantify final-state uncertainty by the basin entropy and the basin boundary entropy. Finally, we recall the Wada property, for the case of three or more escape basins. This property is verified both qualitatively and quantitatively, using a grid approach.
\end{abstract}

\maketitle

\begin{quotation}

In magnetically confined plasmas, the presence of impurities is unavoidable, with great impact on the confinement. In the plasma edge, their presence helps the distribution of the power loss, while in the core it can cause disruptions on the plasma. So understanding and predicting their behavior is necessary for obtaining controlled thermonuclear fusion. The escape of these particles in the plasma edge is directly related to the chaotic transport that arises in particular from electrostatic fluctuations. The trajectories of impurities can be described by the $\mathbf{E}\times\mathbf{B}$ drift motion, which is a hamiltonian dynamics where particle space coordinates are canonically conjugate variables, and chaotic orbits will be able to escape by passing through some exit in this phase space. The set of points with orbits escaping via this exit defines the basin of escape of this exit. In the case of two or more exits, the boundary of basins is usually fractal. Thus fractality is directly related with the chaotic saddle, the invariant non-attractive set of points, formed by the tangle of stable and unstable invariant manifolds, as the unstable manifold represents the escape channels of particles, and the stable manifold traces the boundary of basins. For the case of three exits, the boundary may be shared by three basins, having the topological property called Wada. In this case, the points of different basins are so mixed that the future state of the system is unpredictable. In this work, we explore the escape basins structures, in a two-dimensional symplectic map model for the $\mathbf{E}\times\mathbf{B}$ drift motion, and we quantify the fractality and the uncertainty associated with it, using different methods to understand the transport of these particles. 
\end{quotation}

\section{Introduction}

The understanding of anomalous transport is one of the most important theoretical issues in the quest for controlled thermonuclear fusion \cite{horton2015}. In particular, we are concerned with the chaotic transport of charged particles advected by a turbulent electric field in a magnetized plasma \cite{horton2018}. In the context of the guiding center motion approximation, with the ${\bf E}\times{\bf B}$ drift velocity, this becomes an advection problem described by a low-dimensional Hamiltonian \cite{horton1985}. 

Electrostatic fluctuations are thought to be responsible for turbulent transport in magnetically confined plasmas \cite{flutuacoes}. Such a mechanism has been found to agree with experimental evidence for low plasma pressure \cite{scott}. We limit ourselves to the anomalous transport of trace impurities that are so diluted that their presence does not alter the electric field. The problem becomes analogous to the Lagrangian description of passive scalar advection due to a given stream function, for a two-dimensional incompressible fluid flow \cite{advection,ottino}. Under these assumptions, the ${\bf E}\times{\bf B}$ drift motion of test particles is an exact model for the anomalous transport \cite{ciraolo}. Moreover, the ${\bf E}\times{\bf B}$ drift motion is observed in many magnetized plasma devices \cite{yves2021}, like magnetrons for material processing, many fusion devices, and Hall thrusters in which this term plays a major role in the anomalous transport of particles \cite{yves2020}.

One of the remarkable features of models of ${\bf E}\times{\bf B}$ drift motion is that chaotic particle motion is possible even for regular spatial configurations of the electric field, provided the corresponding Hamiltonian system is time-dependent, and thus non-integrable \cite{horton1985,flutuacoes}. If the flow is time dependent, the advection is chaotic\cite{ottino}, leading to a complex and intricate pattern of advection. Such chaotic motion becomes a non-collisional source of enhanced cross-field particle diffusion, and has been found to yield results many orders of magnitude larger than neoclassical transport \cite{yves2020,pettini,amato}. 

Particle escape in the plasma edge region is an issue directly related to chaotic cross-field transport. For example, if a chaotic orbit connects the plasma outer region and the tokamak inner wall, all particles with initial conditions therein will eventually escape the plasma and hit the tokamak wall. This phenomenon can be harnessed in order to divert particles from the plasma edge into carefully placed collecting plates called divertors \cite{5}.  


A further complication is that the distribution of heat and particle loadings is highly nonuniform, when the particle trajectories related to escape are chaotic. This nonuniformity can be attributed to a geometrical structure underlying chaotic orbits, called homoclinic and heteroclinic tangle, formed by the infinite number of intersections between stable and unstable invariant manifolds emanating from unstable periodic orbits embedded in the chaotic orbit \cite{lai-tel,elton}. Recently it was shown that mode-coupling is enhanced in phase space regions occupied by homoclinic tangles \cite{meirielen}. The unstable manifolds, in particular, represent escape channels for particles in a chaotic orbit, and their geometry influences the spatial distribution of escape patterns \cite{sommerer,evans}. This mechanism is capable to explain qualitatively experimental observations of heat fluxes deposited on divertor plates of tokamaks \cite{wingen}.

In this paper, we explore these ideas to investigate the presence of fractal structures related to particle escape in toroidal devices undergoing chaotic trajectories. Our numerical simulations will be performed using a particle ${\bf E}\times{\bf B}$ drift model in presence of electrostatic fluctuations proposed by Horton {\it et al.} \cite{horton}. The spectrum of electrostatic fluctuations is chosen so as to reduce the dynamics to a two-dimensional, area-preserving map characterizing a non-integrable Hamiltonian system. For values of the physical parameters taken from the Brazilian tokamak TCABR, and using the intensity of the fluctuating electrostatic potential as the tunable parameter, we typically obtain large chaotic orbits extending from the outer portion of the plasma to the tokamak inner wall. As the results can be extended to other machines using an appropriate normalization\cite{horton}, we present a conceptual investigation rather than detailed comparisons with experiments performed in any tokamak.

The phase space of non-integrable Hamiltonian systems is neither entirely regular, nor entirely chaotic. The regular dynamics consists of quasiperiodic orbits lying on tori and periodic orbits, while chaotic orbits fill densely other parts of the energy surface\cite{livro_chaotic}. In order to characterize the phase space, we use the weighted Birkhoff average method\cite{meiss1} to determine whether an initial condition was inside an island or in chaotic sea.  

The most fundamental structure to be studied, in the context of particle escape in chaotic area-filling orbits, is the set of escape basins, which are the sets of particle positions leading to escape through a certain exit. The escape basin boundary coincides with the stable manifold of the chaotic saddle and thus has the same geometrical properties. As a consequence of fractality, we have a sensitive dependence on initial condition, with respect to what region will the chaotic trajectory escape through. Recently, Mathias \textit{et al.}\cite{amanda-phys-a} studied the structures related with the escape of particles through different exits in the boundary of the plasma, caused by two ${\bf E}\times{\bf B}$ drift waves. 

We use a number of quantitative diagnostics for characterizing the fractality of these structures, namely the uncertainty exponent (related to the box-counting dimension) and the corresponding information entropy. Both quantify the final-state uncertainty of the system, i.e. how much an improvement of the precision in the determination of an initial condition is reflected in the uncertainty of finding what exit will the corresponding trajectory escape through. We also identify the Wada property, which is typical for three or more escape regions, and characterizing an extreme form of fractal behavior. A natural question is how particle collisions affect these fractal structures. To answer these questions, we use a simple numerical model where collisions can be regarded as a noisy component in the ${\bf E}\times{\bf B}$ drift equations.

The rest of this article is organized as follows: in Section II, we outline the symplectic (area-preserving) map describing chaotic advection of test particles in the ${\bf E}\times{\bf B}$ drift motion caused by a radial equilibrium electric field plus electrostatic fluctuations. Section III is devoted to a detailed discussion of the radial profiles of the equilibrium safety factor of the magnetic field, the electric field, and the toroidal velocity of the plasma. Section IV uses the concept of weighted Birkhoff averages and their long-time convergence as a diagnostic of chaotic motion. Section V discusses escape basins and their underlying mathematical structure. Section VI deals with the numerical characterization of fractal structures using the uncertainty exponents and basin entropies. The Wada property and its quantitative characterization are discussed in Section VII. In Section VIII we discuss the effect of collisions on the fractal structures. The last Section is devoted to our Conclusions. 

\section{Symplectic map for drift motion}

Let us denote by $a$ and $R_0$ the minor and major plasma radius, respectively, in a tokamak. In the following, we will describe the particle position using local coordinates $(r,\theta,\varphi)$, where $r$ is measured from the minor axis, $\theta$ is the poloidal angle, and $\varphi$ the toroidal angle. We assume a large aspect ratio approximation, ($\epsilon = a/R_0 \ll 1$), such that the equilibrium magnetic field is $\mathbf{B} =\left(0,B_\theta(r),B_\varphi\right)$, where $B_\varphi$ and $B_\theta$ are the toroidal and poloidal components, respectively. 

Moreover, since $B_\theta \sim \epsilon B_\varphi$, we have $B \approx B_\varphi \gg B_\theta$ and thus consider $B$ as a uniform field. In this approximation, the magnetic (flux) surfaces are nested tori with circular cross sections, with a radial profile for the corresponding safety factor
\begin{equation}
\label{qr}
    q(r) = \frac{r B}{R_0 B_\theta(r)}.
\end{equation}
In this work, we consider two kinds of electrostatic fields: (i) an external and time-independent electric field in the radial direction, and (ii) the time-dependent field related to the drift instabilities, in the form
\begin{equation}
    \label{efield}
    {\bf E} = \bar{E}_r(r) {\hat{\bf r}} - \nabla{\tilde\phi}(r,\theta,\varphi;t),
\end{equation}
where $\tilde{\phi}$ is the electric potential of the drift instabilities. Under these conditions, the guiding-center motion can be thought as a superposition of a passive advection along the magnetic field lines, with velocity $v_\parallel$, and an ${\bf E}\times{\bf B}$ drift velocity. The resulting equation of motion for the guiding-center is thus   
\begin{equation}
   \label{eqm}
    \frac{d{\bf r}}{dt} = v_\parallel \frac{\textbf{B}}{B} + \frac{\textbf{E}\times\textbf{B}}{B^2},
\end{equation}
which gives the components
\begin{align}
    \label{eqr}
    \frac{dr}{dt} & = - \frac{1}{rB} 
    \frac{\partial{\tilde\phi}}{\partial \theta}, \\
    \label{eqt}
    \frac{d\theta}{dt} & = \frac{v_\parallel(r)}{R_0q(r)} - \frac{\bar{E}_r(r)}{rB} + \frac{1}{rB} \frac{\partial{\tilde\phi}}{\partial r}, \\
    \label{eqf}
    \frac{d\varphi}{dt} & = \frac{v_\parallel(r)}{R_0}.
\end{align}

The functions $v_\parallel,\,E_r$ and $q$ will be given in the next section. The electric potential related to the drift instabilities is assumed to exhibit a broad spectrum of frequencies $\omega_n = n \omega_0$ and wave vectors, characterized by a Fourier expansion in the general form \cite{horton}
\begin{equation}
\label{eq:spectrum}
\tilde{\phi}(r,\theta,\varphi;t) = \sum_{m,\ell,n}  \phi_{m, \ell, n} (r) \cos(m\theta-\ell\varphi -n\omega_0 t + \nu_{m, \ell, n}),
\end{equation}
where the coefficients $\phi_{m,\ell,n}$ depend, in general, on the radius $r$ and time but, for simplicity, we assume them constant over the plasma region of interest in this work, and following Horton \textit{et al}\cite{horton}, we take $\nu_{m, \ell, n} = 0$ for all waves. Moreover, we retain only the dominant Fourier mode in Eq. (\ref{eq:spectrum}), with harmonics of the lowest frequency $\omega_0$, and fixed poloidal and toroidal mode numbers $m = M$ and $\ell = L$, respectively, with the same amplitude $\phi_{M,L} = \phi$ for all harmonics, such that 
\begin{equation}
    \label{eq:total.spectrum}
    \tilde{\phi}(\theta,\varphi;t) = 2\pi\phi\cos{(M\theta-L\varphi)} \sum_n \delta(\omega_0 t - 2\pi n)
\end{equation}
where we used the formulas
\begin{align}
    \label{pois}
    \sum_{n=-\infty}^{+\infty} \cos(n\omega_0 t) & = 2\pi \sum_n \delta(\omega_0 t - 2\pi n), \\
    \label{poiss}
    \sum_{n=-\infty}^{+\infty} \sin(n\omega_0 t) & = 0.
\end{align}

The drift motion of guiding centers is a Hamiltonian system, with canonical equations
\begin{equation}
    \label{hh}
    \frac{dI}{dt} = - \frac{\partial H}{\partial\Psi}, \qquad 
    \frac{d\Psi}{dt} = \frac{\partial H}{\partial I},  
\end{equation}
where we define action and angle variables by $I=(r/a)^2$ and $\Psi= M\theta - L\varphi$, respectively. Making this transformation and using equation (\ref{qr}), reduces the system (\ref{eqr})-(\ref{eqf}) to the form
\begin{align}
    \label{eqr1}
    \frac{dI}{dt} & = \frac{4\pi M \phi}{a^2B} \, \sin\Psi \sum_{n} \delta(\omega_0 t - 2\pi n), \\
    \label{eqt1}
    \frac{d\Psi}{dt} & = \frac{v_\parallel(I)}{R_0 q(I)} \, (M - q(I) L) - \frac{M \bar{E}_r(I)}{aB\sqrt{I}}.
\end{align}

We define discrete variables by considering a stroboscopic sampling of the action-angle variables at integer multiples of the characteristic period
\begin{align}
    \label{stroboI}
    I_n & = \lim_{\eta \searrow 0}  I\left( t = \frac{2\pi n}{\omega_0} - \eta \right), \\
    \label{strobo2}
    \Psi_n & = \Psi\left(t = \frac{2\pi n}{\omega_0}\right),
\end{align}
leading to the two-dimensional Poincaré map
\begin{align}
    \nonumber
    I_{n+1} & = I_n + \frac{4\pi M \phi}{a^2B\omega_0} \sin\Psi_n, \\
    \nonumber
    \Psi_{n+1} & = \Psi_n + \frac{2\pi v_\parallel(I_{n+1})}{\omega_0 R_0} \frac{M - L q(I_{n+1})}{q(I_{n+1})} \\
     \nonumber
    & - \frac{2\pi M}{aB\omega_0} \frac{\bar{E}_r(I_{n+1})}{\sqrt{I_{n+1}}}.
\end{align}

We applied a normalization to the quantities $B,\, a,\, \omega_0,\, \bar{E}_r,\, \phi, \, R_0$, the minor plasma radius is divided by $a_0=0.18\,{\textrm{cm}}$, so that the normalized value is $a'=1$, in the same manner the toroidal magnetic field is divided by $B_0=1.1\,{\textrm{T}}$, giving the normalized value $B'=1.0$. We chose a normalization for the eletric field, in order for the normalized value to be equal to unity in the plasma edge, $E_r=\bar{E}_r/E_0$. The normalization factors for the other quantities are given in terms of $a_0,\,B_0$ and $E_0$, namely, velocity $v_0=E_0/B_0$, time $t_0=a_0/v_0$, so $\omega=\omega_0t_0$, electrical potential $\phi_0=a_0E_0$, and $R_0'=R_0/a_0$. The normalized map equations are
\begin{align}
    \label{map1}
    I_{n+1} & = I_n + \frac{4\pi M \phi}{\omega} \sin\Psi_n, \\
    \nonumber
    \Psi_{n+1} & = \Psi_n + \frac{2\pi a_0 v_\parallel(I_{n+1})}{\omega R_0} \frac{M - L q(I_{n+1})}{q(I_{n+1})} \\
     \label{map2}
    & - \frac{2\pi M}{\omega} \frac{\bar{E}_r(I_{n+1})}{\sqrt{I_{n+1}}}.
\end{align}

\section{Radial profiles}
 
The map defined by (\ref{map1})-(\ref{map2}) is area-preserving in the phase plane $(I,\Psi)$ corresponding to the Poincaré surface of section obtained by using (\ref{stroboI})-(\ref{strobo2}). It is important to emphasize that the Poincaré surface of section we deal with is in fact a stroboscopic sampling of the action and angle variables, rather than a fixed plane in space, like at $\varphi = $ const. Such a description would be possible by numerically solving the differential equations of motion (\ref{eqr})-(\ref{eqf}) and considering the intersections of the particle trajectory with a fixed plane. Hence in the present work, we will be interested in analyzing the escape in the phase plane of action-angle variables. 

In order to investigate the dynamics generated by iterating the map (\ref{map1})-(\ref{map2}), we have first to give analytical expressions for three radial profiles: the safety factor $q(I)$, the radial electric field $E_r(I)$, and the toroidal velocity $v_\parallel(I)$. In this work, we use parameters\cite{nascimento} of the TCABR tokamak, operating at the Physics Institute of São Paulo University (Brazil), listed in Table \ref{tab:parameters}.

\begin{table}[htb]
\begin{center}
\begin{tabular}{|c|c|c|}
\hline 
parameter & symbol & value  \\
\hline 
\hline 
minor radius & $a$ & $0.180~\mathrm{m}$ \\
\hline
major radius & $R_0$ & $0.615~\mathrm{m}$ \\
\hline
toroidal field & $B_\varphi$ & $1.1~\mathrm{T}$ \\
\hline
plasma current & $I_p$ & $100~\mathrm{kA}$ \\
\hline
central electron temperature & $T_{e0}$ & $400~\mathrm{eV}$ \\
\hline
central electron density & $n_{e0}$ & $3.0 \times 10^{19}~\mathrm{m^{-3}}$ \\
\hline
pulse duration & $\tau_p$ & $120~\mathrm{ms}$ \\
\hline
\end{tabular}
\end{center}
\caption{Main parameters of the TCABR tokamak \cite{nascimento}.}
\label{tab:parameters}
\end{table}

We used $M=15$ and $L=6$ as the main poloidal and toroidal modes \cite{marcus2}, typical numbers in the wave spectrum at the tokamak plasma edge \cite{horton}. The normalized fundamental angular frequency is $\omega=16.36$.

\begin{figure*}
    \centering
    \subfloat(a){\includegraphics[height=2.in]{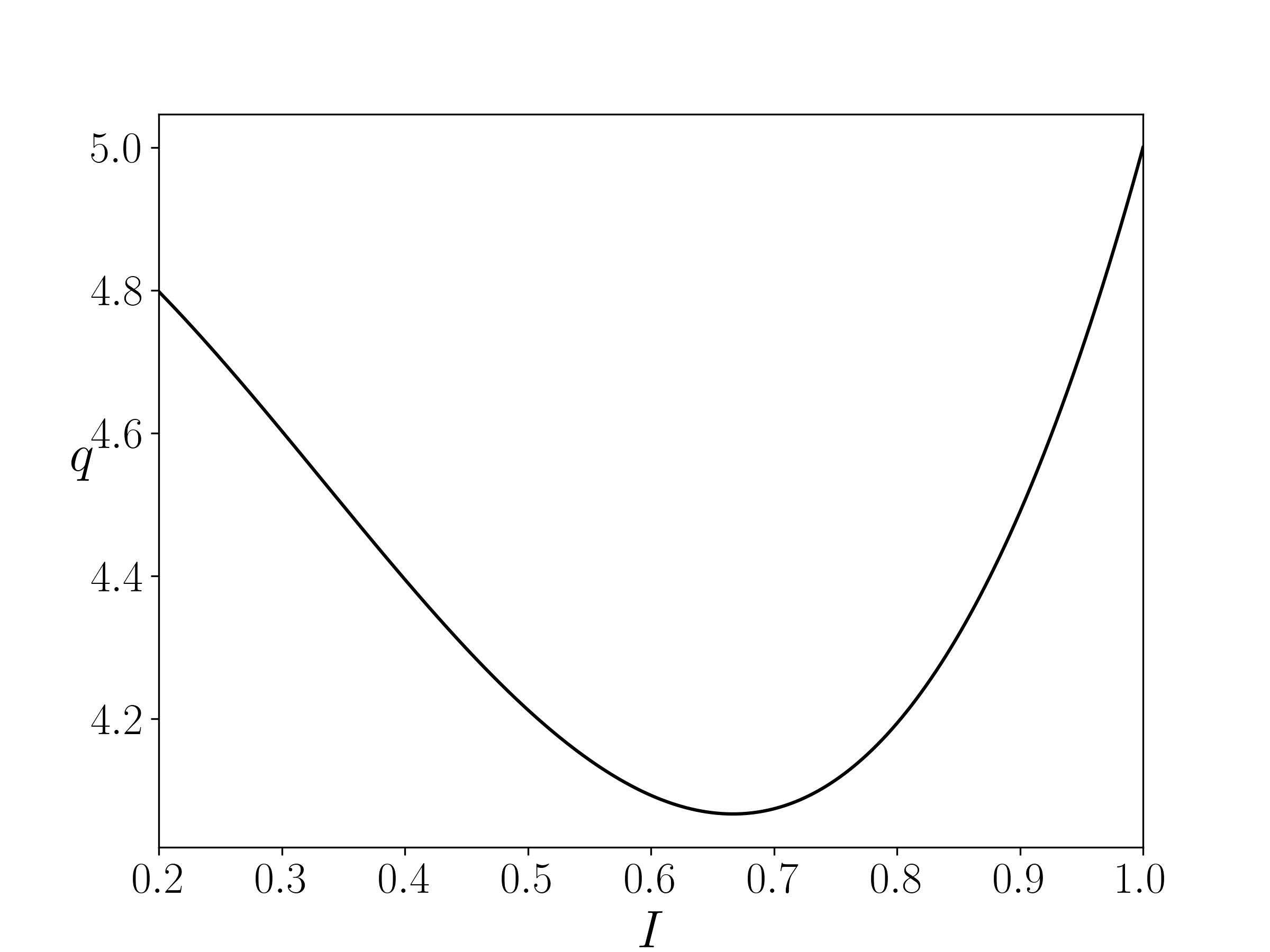}}
    \subfloat(b){\includegraphics[height=2.in]{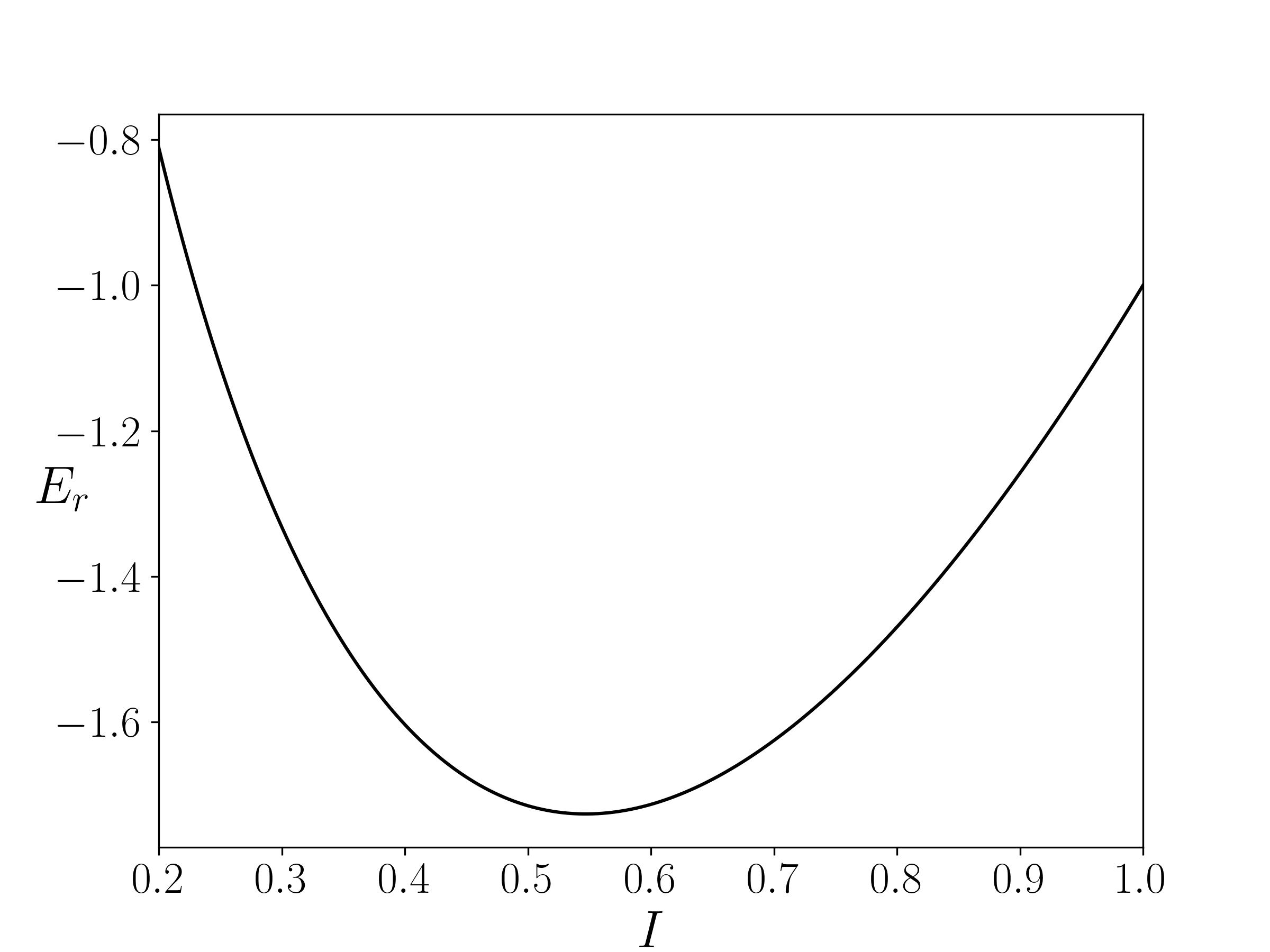}}
    \subfloat(c){\includegraphics[height=2.in]{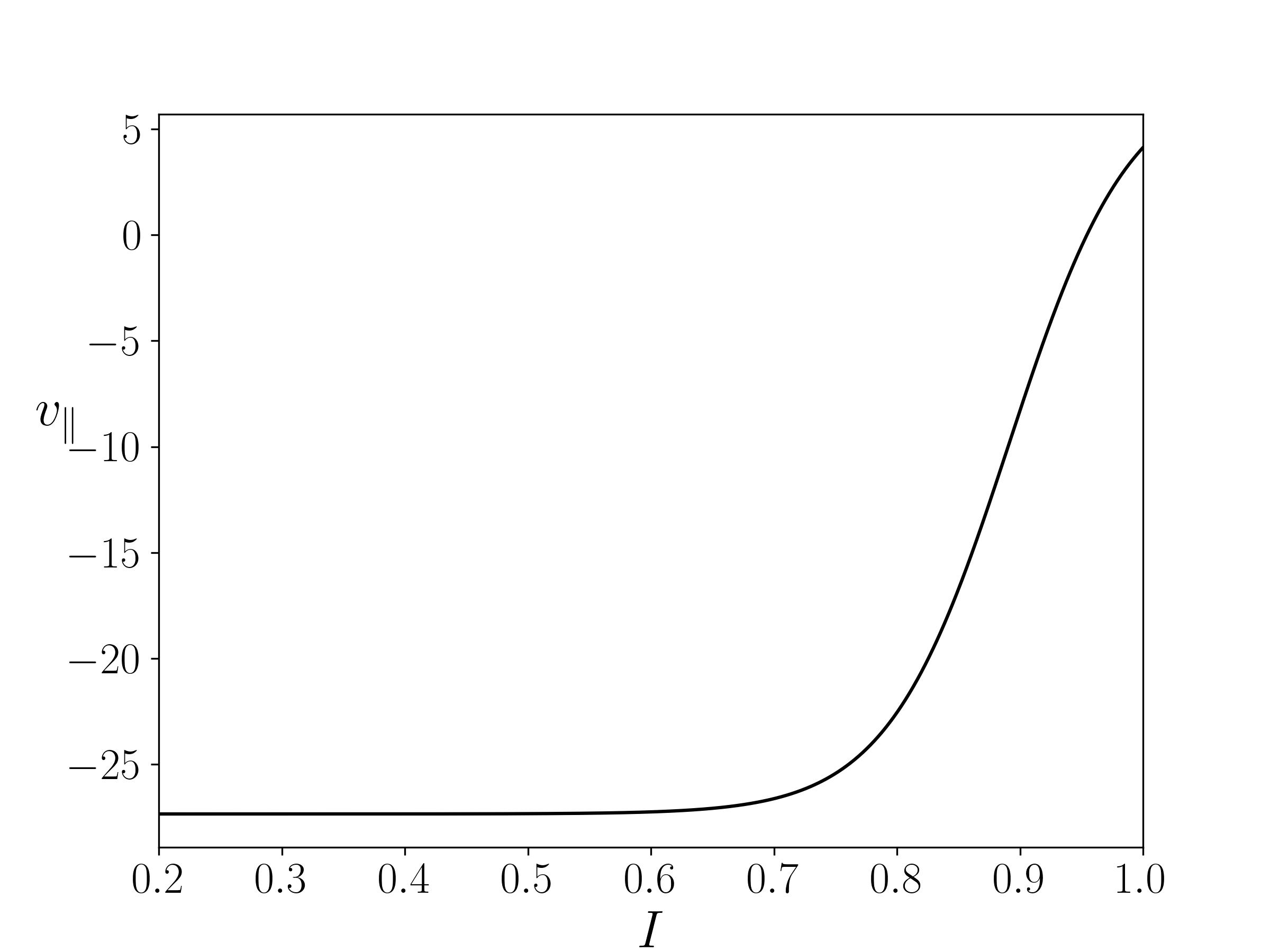}}
    \caption{Radial profiles in terms of the action variable $I  = r^2/a^2$ for the quantities: (a) safety factor, (b) equilibrium electric field, and (c) toroidal plasma velocity.}
    \label{fig:profiles}
\end{figure*}

Non-monotonic safety factors generate negative shear regions in the plasma, which improves the plasma confinement quality. There is a significant reduction of turbulent transport by using this type of safety factor \cite{levinton-1995,strait-1995}. The radial safety factor we considered, in terms of the action variable $I = r^2/a^2$, is 
\begin{equation}
    \label{eq:safety.profile}
    q(I) = 5.0 - 6.3 \, I^2 + 6.3 \, I^3,
\end{equation} 
so that $q(I=1) = 5.0$ at the plasma edge, which is consistent with measurements of plasma current, electron density and temperature in TCABR tokamak [FIG. \ref{fig:profiles}(a)]. 

Turbulent particle fluxes in H-mode tokamak discharges are reduced by the presence of a radial electric field with negative shear \cite{viezzer,hidalgo}, generating a shearless transport barrier \cite{marcus,marcus2} that is compatible with the reduction of the turbulent fluxes. We adopt the profile
\begin{equation}
\label{eq:electric.field.profile}
E_r(I) = 10.7 \, I - 15.8 \, \sqrt{I} + 4.13,
\end{equation} 
so as to yield a local minimum in the desired plasma region \cite{nascimento} [FIG.  \ref{fig:profiles}(b)]. It should be noted that the model functions (\ref{eq:safety.profile})-(\ref{eq:electric.field.profile}) may result in ill-defined physics at the singularity of polar coordinates where $I=0$. Nevertheless, it is worth mentioning that the numerical trajectories we will be computing always remain in the range of $I\geq 0.2$, rendering the actual model irrelevant near $I=0$.

We take into account the plasma rotation by considering a non-monotonic profile for the toroidal plasma velocity, which is related with shearless barriers \cite{ferro}. Spectroscopic techniques have been used  to measure toroidal plasma rotation velocities in TCABR discharges, giving values about $4.0~\mathrm{km/s}$ at the plasma edge \cite{nascimento}. A normalized parallel velocity profile, used in this work and consistent with TCABR observations, is given\cite{ferro} [FIG. \ref{fig:profiles}(c)] by
\begin{equation}
    \label{eq:toroidal.velocity}
    v_\parallel(I) = - 9.867 + 17.47 \, \tanh(10.1 \, I - 9.00).
\end{equation} 

In our numerical simulations, we integrate the map (\ref{map1})-(\ref{map2}) using the profiles for the equilibrium safety factor, radial electric field, and toroidal velocities given by Eqs. (\ref{eq:safety.profile}), (\ref{eq:electric.field.profile}), and (\ref{eq:toroidal.velocity}), respectively. We keep all parameters fixed and choose the amplitude of the main electrostatic mode $\phi$ as the variable parameter. Proceeding in this way, we can evaluate the qualitative effects of increasing perturbation strength on the orbit structure generated by the map (\ref{map1})-(\ref{map2}). 

\begin{figure*}
    \centering
    \subfloat(a){\includegraphics[height=2.5in]{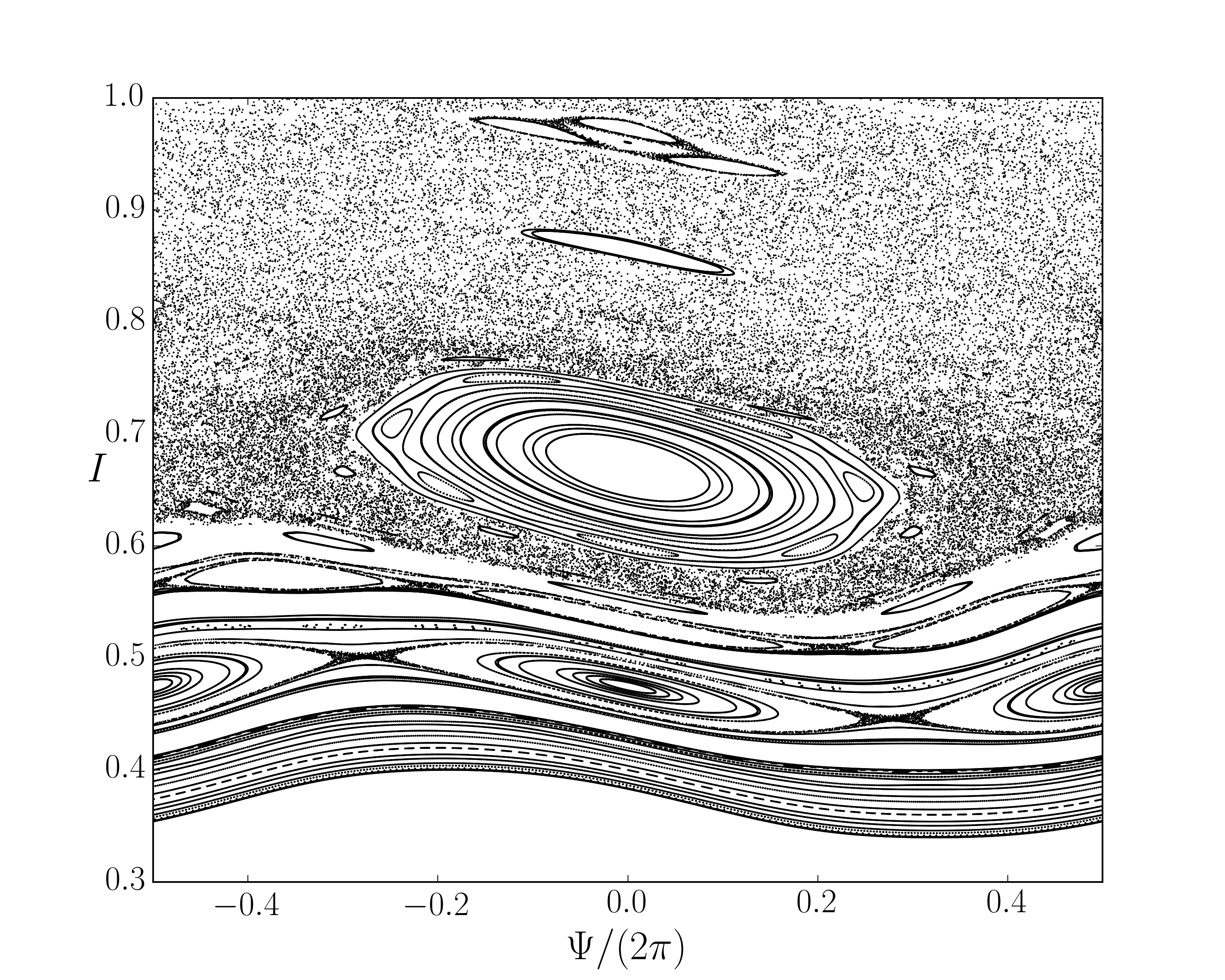}}
    \subfloat(b){\includegraphics[height=2.5in]{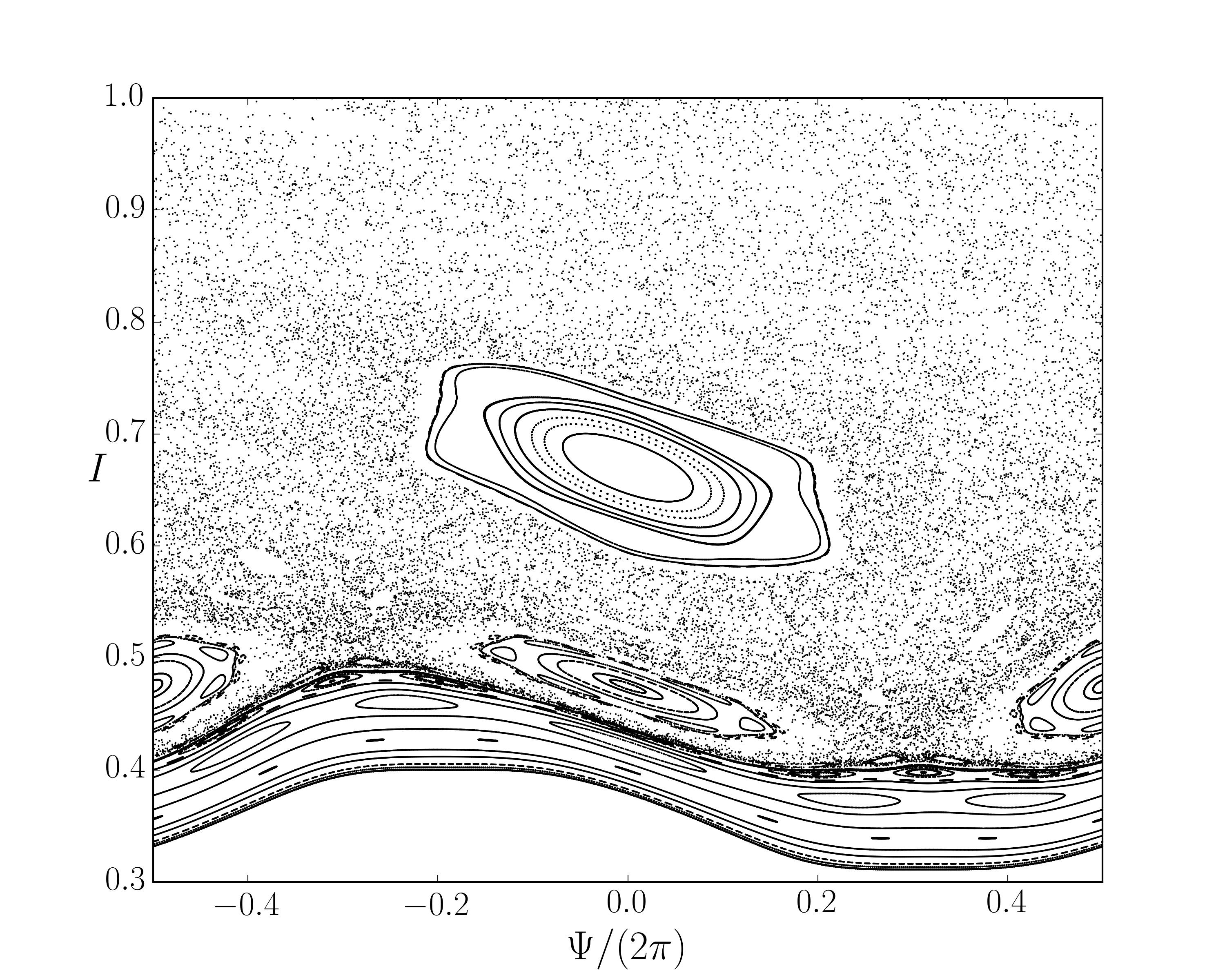}}
    \subfloat(c){\includegraphics[height=2.5in]{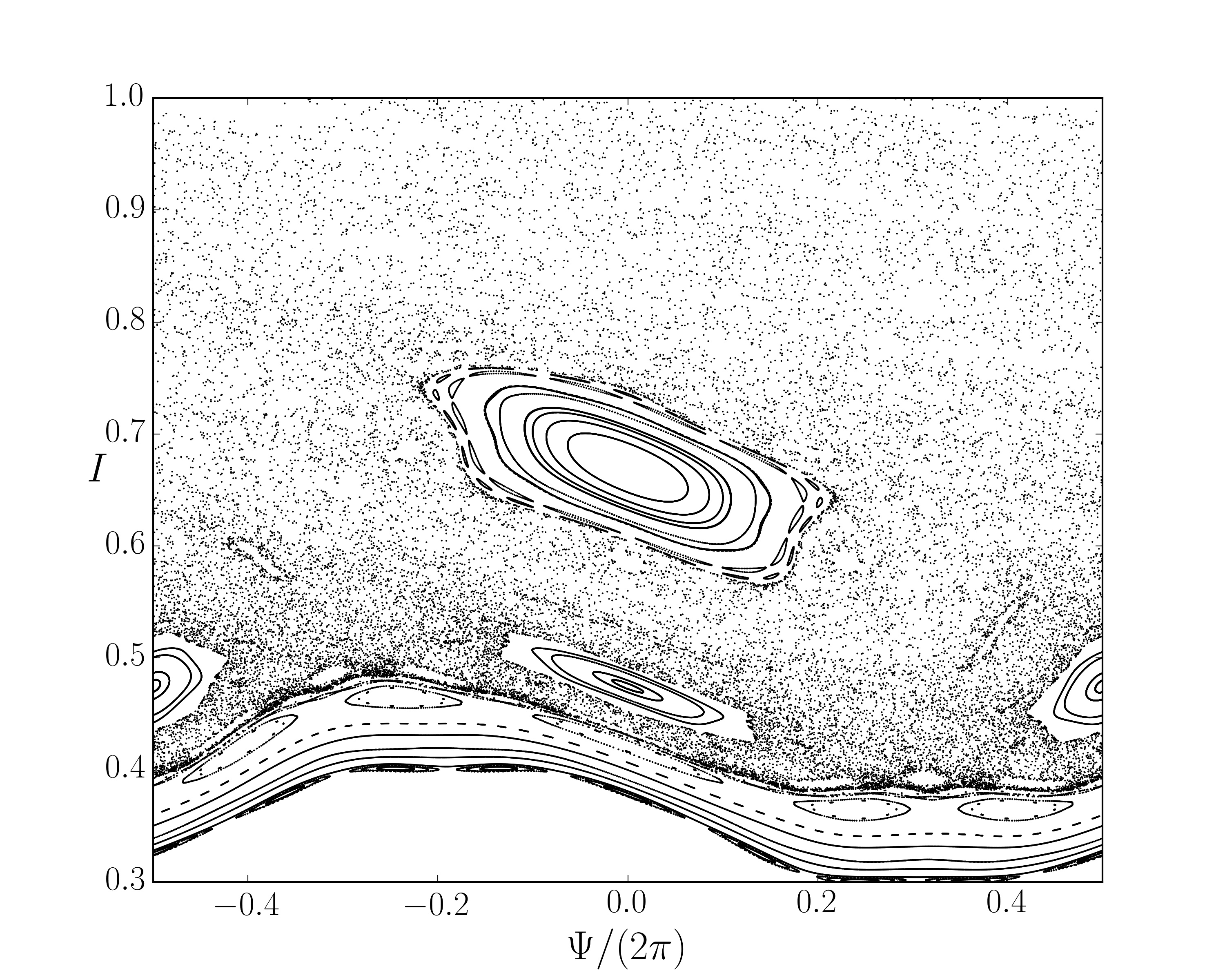}}
    \subfloat(d){\includegraphics[height=2.5in]{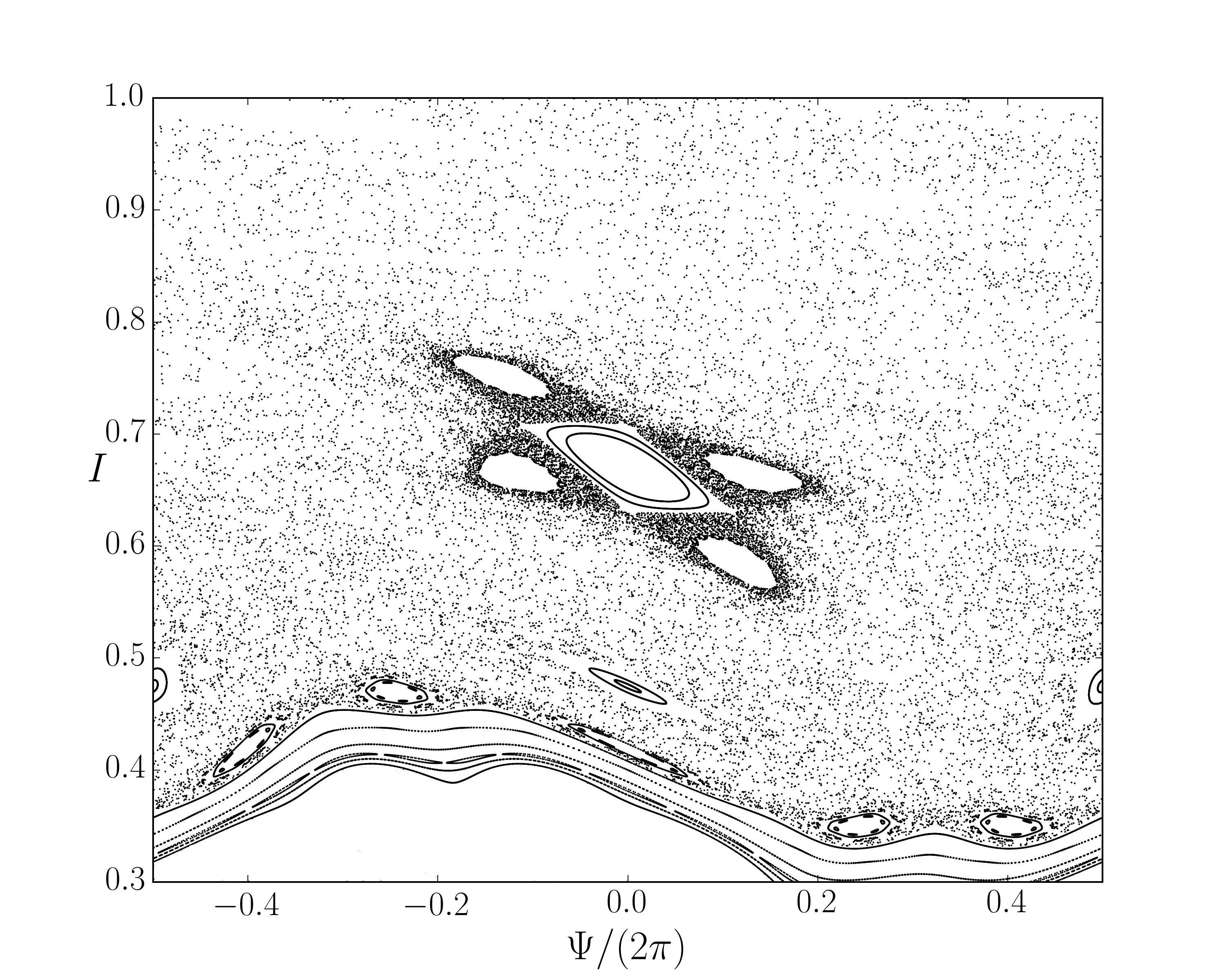}}
    \caption{Phase portraits of the map (\ref{map1})-(\ref{map2}) for the following values of the perturbation amplitude $\phi$: (a) $4.92\times 10^{-3}$, (b) $7.65\times 10^{-3}$, (c) $8.74\times 10^{-3}$, and (d) $10.38\times 10^{-3}$.}
    \label{fig:phase}
\end{figure*}

Figure \ref{fig:phase} depicts some phase portraits of the map, using rectangular coordinates for $I_n$ and $\Psi_n/(2\pi)$, for ease of visualization. For a relatively small value of $\phi$, we have a divided phase space consisting of an outer large chaotic sea, with remnants of periodic islands embedded in it, and an inner structure of invariant tori and island chains comprising the plasma core. The large chaotic sea intercepts the plasma boundary at the radial distance corresponding to $I = 1.0$, in such a way that an initial condition placed within the chaotic orbit will eventually escape the plasma through that boundary. This is thus an open Hamiltonian system. On the other hand, those initial conditions placed in the inner region are not expected to escape due to the invariant tori which act as dikes, preventing large-scale chaotic transport [FIG. \ref{fig:phase}(a)]. As the perturbation strength increases, the outer chaotic region is enlarged by engulfing some of the nearby invariant tori and island chains [FIG. \ref{fig:phase}(b)-(c)]. For large $\phi$, the chaotic region encompasses virtually all the region formerly occupied by the plasma column [FIG. \ref{fig:phase}(d)].

\section{Weighted Birkhoff averages}

The phase space of a non-integrable Hamiltonian system exhibits basically three types of trajectories: chaotic orbits, periodic island chains, and invariant tori. Although, in principle, these types of trajectories could be straightforwardly identified by known methods (Lyapunov exponents, rotation numbers, etc.), in practice there is a complication introduced by the self-similar hierarchical structure, in which islands and chaotic orbits are mixed together in arbitrarily fine scales. This hierarchical structure is the ultimate reason why it is difficult to estimate with accuracy the destruction of a given spanning invariant torus (a curve, in two dimensions). A major advance in this direction has been the concept of weighted Birkhoff averages\cite{meiss1, meiss2, matheus}. 

Let ${\bf M}$ represent our two-dimensional symplectic map describing drift trajectories of passive particles in a ${\bf E} \times {\bf B}$-flow. For brevity, let us denote the two-dimensional phase space vector at discrete time $n$: ${\bf v}_n = (I_n,\Psi_n)$, in such a way that (\ref{map1})-(\ref{map2}) are written in the compact form
\[
{\bf v}_{n+1} = {\bf M}({\bf v}_n),
\]
that can be iterated $n$ times, from an initial condition ${\bf v}_0$
\[
{\bf v}_{n} = {\bf M}^n({\bf v}_0).
\]
The Birkhoff average of some function $f({\bf v})$ along this trajectory in phase space is defined as 
\begin{equation}
\label{ba}
B_N(f)({\bf v}_0) = \frac{1}{N} \sum_{n=0}^{N-1} f \circ {\bf M}^n({\bf v}_0).
\end{equation}

According to the Birkhoff ergodic theorem, time averages of the function $f$ along the trajectory (i.e. Birkhoff averages) converge to the phase space averages as $N \rightarrow \infty$
\begin{equation}
\label{bet}
\frac{1}{N} \sum_{n=0}^{N-1} f \circ {\bf M}^n({\bf v}_0) \rightarrow \int f d\mu,
\end{equation}
where $\mu$ is an invariant probability measure. A noteworthy fact about Birkhoff’s ergodic theorem is that the convergence to the phase space average can be very slow. For example, for a quasiperiodic orbit in an invariant torus the convergence rate of (\ref{bet}) scales as $N^{-1}$, whereas for chaotic orbits this rate varies as $N^{-1/2}$. The reason for this slow convergence is the lack of smoothness caused by the two ends of a finite orbit segment, also called edge effect. 

In order to circumvent this smoothness problem there has been proposed a weighted Birkhoff average
\begin{equation}
\label{wba}
WB_N(f)({\bf v}_0) = \sum_{n=0}^{N-1} w_{n,N} f \circ {\bf M}^n({\bf v}_0),
\end{equation}
where we define weights
\begin{equation}
\label{weights}
w_{n,N} = \frac{g(n/N)}{\sum_{n=0}^{N-1} g(n/N)},
\end{equation}
where an exponential bump function
\begin{equation}
\label{bump}
g(z) = \begin{cases}
          \exp\{-{\lbrack z(1-z)\rbrack}^{-1} \} & \text{if $0 < z < 1$} \\
          0 & \text{otherwise}
       \end{cases}
\end{equation}
was chosen to smooth the behavior at the ends of a finite-time trajectory. Indeed, Eq. (\ref{bump}) converges to zero with infinite smoothness at $z = 0$ and $z = 1$. Since the weights vanish flatly at these end points, the smoothness of the original orbit is preserved. 

It has been proved\cite{yorke-dig} that, if the functions $g$, $f$ and the map ${\bf M}$ are infinitely differentiable then the convergence in the Birkhoff ergodic theory is super-polynomial, i.e. for all $m \geq 1$ there exists $C_m$ such that 
\begin{equation}
\label{super}
\left\vert WB_N(f)({\bf v}_0) - \int f d\mu \right\vert \le C_m N^{-m},
\end{equation}

This weight method does not improve the convergence of the Birkhoff's average for chaotic orbits. Interestingly enough, even though the constant $C_m$ depends generally on the function $f$, the speed and accuracy of the convergence of the weighted Birkhoff averages are independent on the choice of $f$\cite{das-1}. Hence we can use a simple function like $f = \cos\Psi$.

Given the difference of convergence for chaotic and quasiperiodic orbits, weighted Birkhoff averages can be used to distinguish among chaotic and regular orbits in the following way: we compute the first $2N$ iterations of the map ${\bf M}$. Then we compare the value of $WB_N(f)({\bf v}_0)$ along the first $N$ map iterations with $WB_N(f)({\bf v}_{N+1})$ along the next $N$ iterations. In the limit $T\rightarrow\infty$, these values are equal and we determine the convergence rate by computing the number of zeroes after the decimal point of the difference between both weighted averages, namely
\begin{equation}
\label{dig}
{\mathrm{dig}} = - \log_{10} \left\vert WB_N(f)({\bf v}_0) - WB_N(f)({\bf v}_{N+1}) \right\vert.
\end{equation}
If ${\mathrm{dig}}$  is a large number, the convergence is fast enough and the orbit is regular. Otherwise (small ${\mathrm{dig}}$) the orbit is chaotic. Care must be taken, however, since the weighted Birkhoff averages are not likely to improve the convergence of chaotic orbits. Hence we cannot compare the values of ${\mathrm{dig}}$ of two chaotic orbits to say what is more chaotic (in the Lyapunov exponent sense). 

\begin{figure*}
    \centering
    \subfloat(a){\includegraphics[height=2.5in]{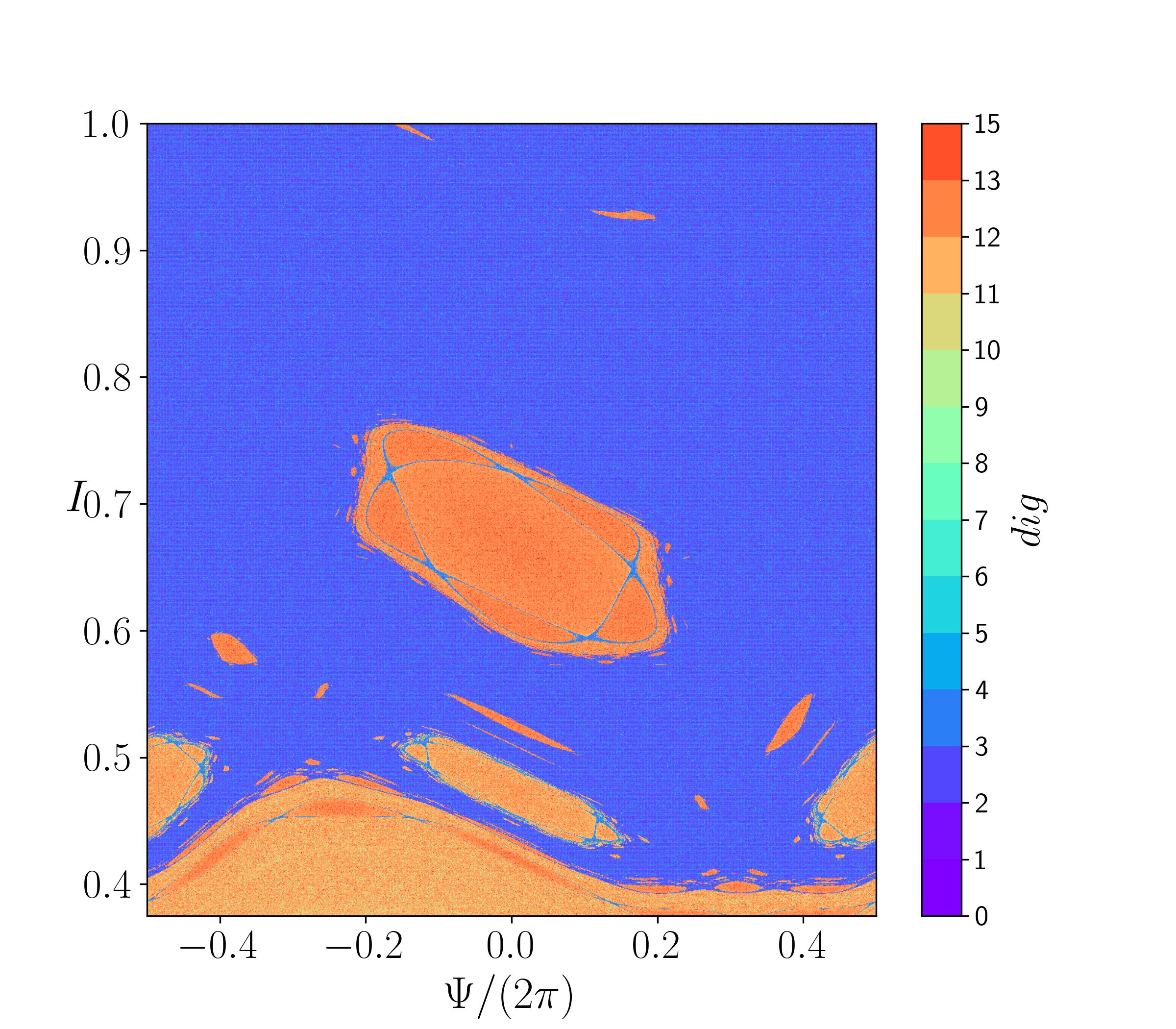}}
    \subfloat(b){\includegraphics[height=2.5in]{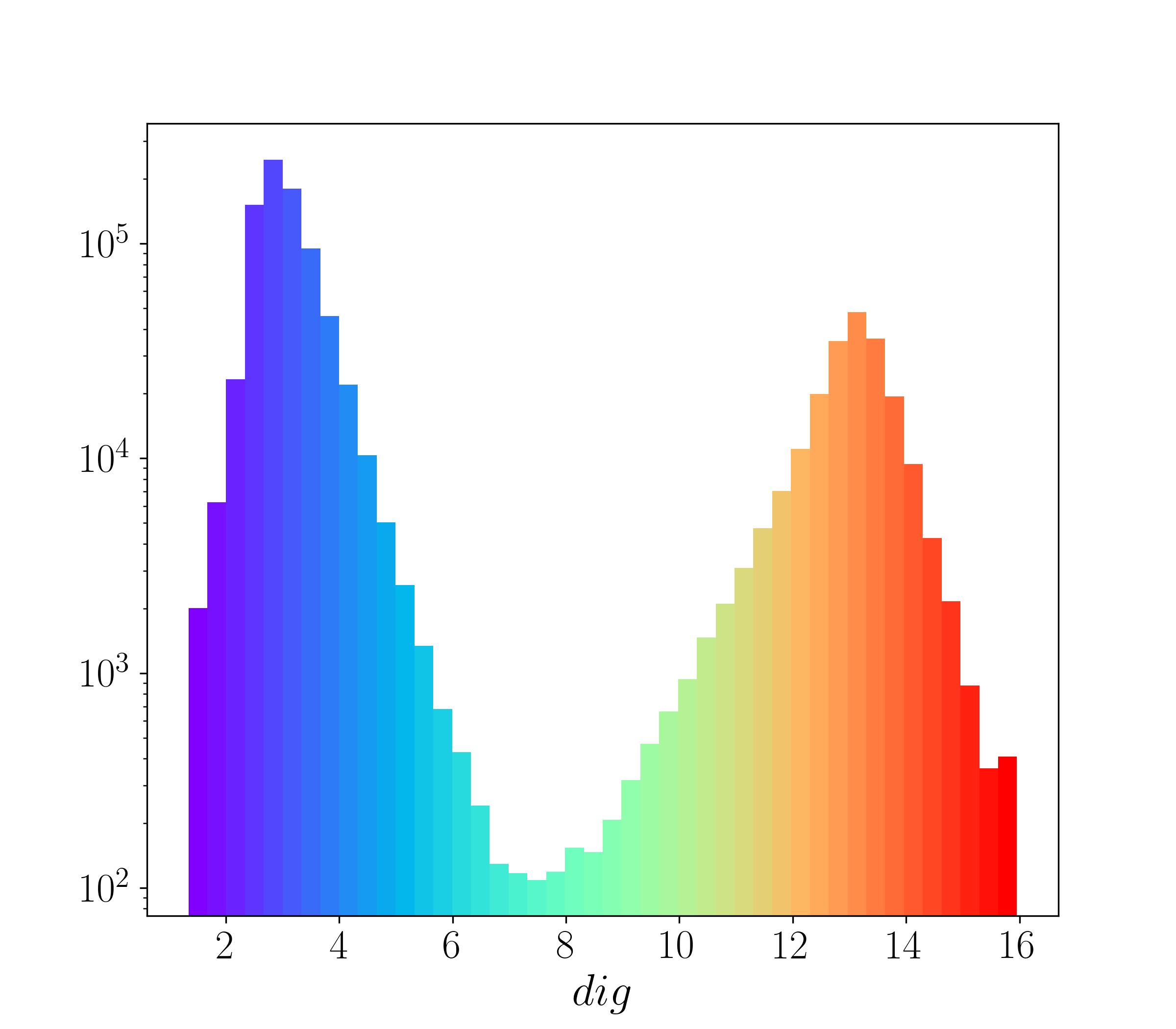}}
    \caption{(a) The number of zeros ${\mathrm{dig}}$ after the decimal point for the map (\ref{map1}-\ref{map2}), and control parameter $\phi = 7.65\times 10^{-3}$; (b) frequency histogram (semi-log scale) of ${\mathrm{dig}}$. The color palette of the histograms matches the color bar of (a).}
    \label{fig:dig}
\end{figure*}

We show, in Fig. \ref{fig:dig}(a), the values of ${\mathrm{dig}}$  for a grid of $1000\times 1000$ initial conditions  superimposed in a phase space rectangle with sides $0.375 < I < 1.0$ and $-0.5 < \Psi/(2\pi) < 0.5$. Each initial condition at the center of grid cells was iterated $2\times 10^6$ times. We used a color palette such that bluish regions correspond to chaotic orbits, whereas reddish region to quasiperiodic orbits. The stronger shades of blue, however, indicate that the density of chaotic points is higher. In the region around $I\sim 0.4$, there are quasiperiodic orbits (KAM curves) that acts as barriers and the chaotic regions (bluish regions) do not reach the plasma core. Therefore, the region of interest to study the escape through leaks in the plasma edge is $0.3\leq I\leq 1.0$. The frequency histogram, Fig. \ref{fig:dig}(b), shows that the regular orbits correspond to the distribution centered around ${\mathrm{dig}}\sim 13$ while the chaotic orbits are centered around ${\mathrm{dig}} \sim 2.8$.


\section{Escape basins}

In this work, we focus on the dynamics of test particles, i.e. charged particles which are passively advected by the drift flow generated by the combined effects of crossed electric and magnetic fields. Such a particle can escape the tokamak by hitting some boundary surface, like that of a divertor plate, similar to those used to mitigate the plasma-wall interactions through exhaustion of particles escaping along a chaotic orbit near a plasma separatrix \cite{punjabi}. 

Instead of investigating directly this type of escape, we will open the dynamical system given by the map equations (\ref{map1})-(\ref{map2}) by considering that the particles are able to escape by one or more exits in the $I \times \Psi$ phase plane \cite{leaks,viana-sanjuan-2007}. Accordingly, we will consider two exits placed at the position $I=1.0$: one from $-\pi\leq\Psi<0.0$, denoted by $L$, and the second exit $0.0\leq\Psi\leq \pi$, denoted by $R$. 

Let us consider an initial condition $(I_0,\Psi_0)$. For each iteration of the map (\ref{map1})-(\ref{map2}), we make the following test: if $I_n \leq 1.0$, we continue iterating, otherwise we stop iterating and consider the value of $\Psi_n$. If $I_n > 1.0$ for some $n \ge 1$ and $-\pi\leq\Psi_n<0.0$, we consider an escape through exit $L$, otherwise through $R$. The sets of initial conditions, for which there is a value of $I_n > 1.0$ ($n \ge 1$) indicating an escape through exits $L$ and $R$, form their corresponding basins of escape, denoted ${\cal B}(L)$ and ${\cal B}(R)$, respectively.

\begin{figure*}
    \centering
    \subfloat(a){\includegraphics[height=2.5in]{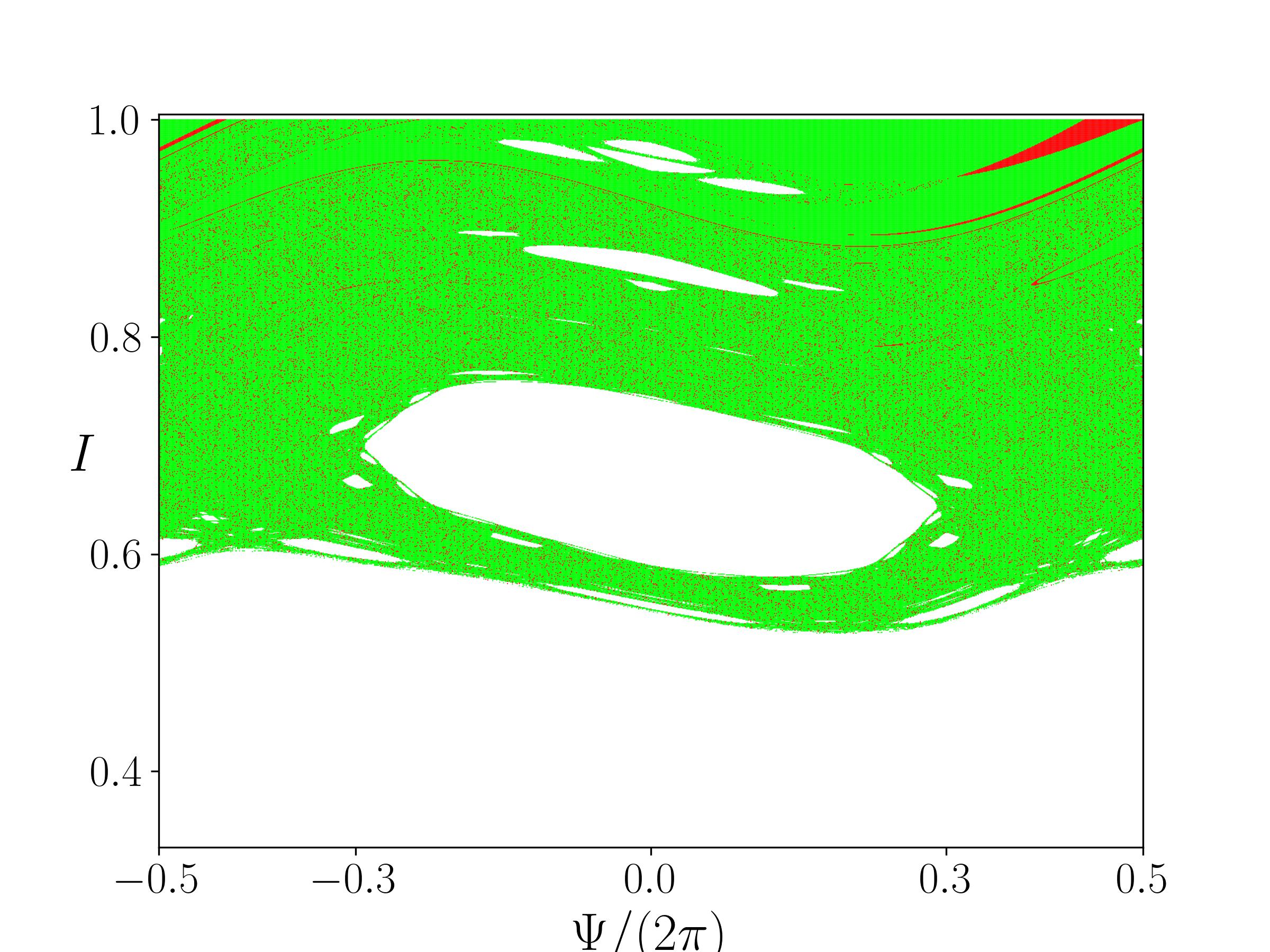}}
    \subfloat(b){\includegraphics[height=2.5in]{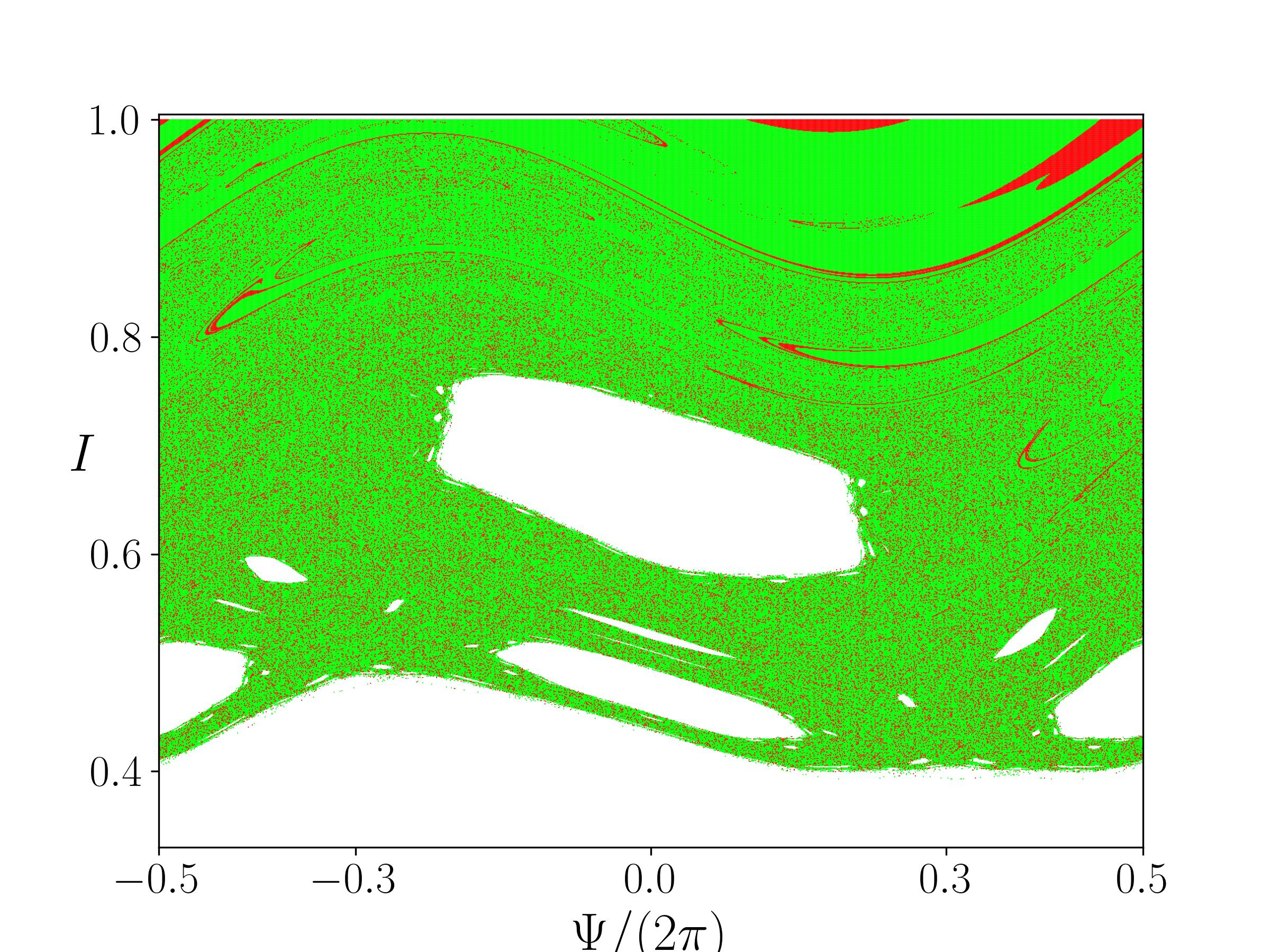}}
    \subfloat(c){\includegraphics[height=2.5in]{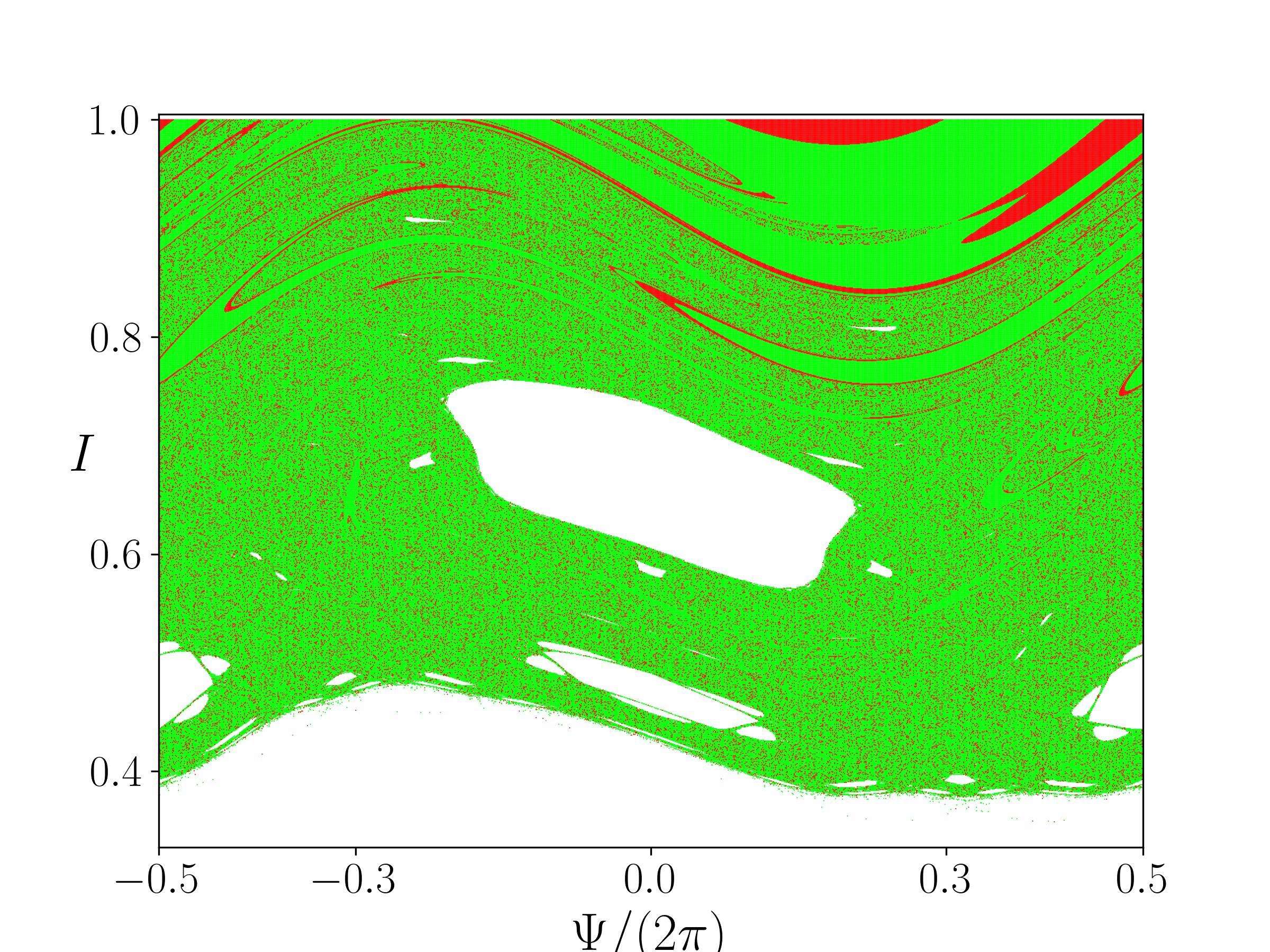}}
    \subfloat(d){\includegraphics[height=2.5in]{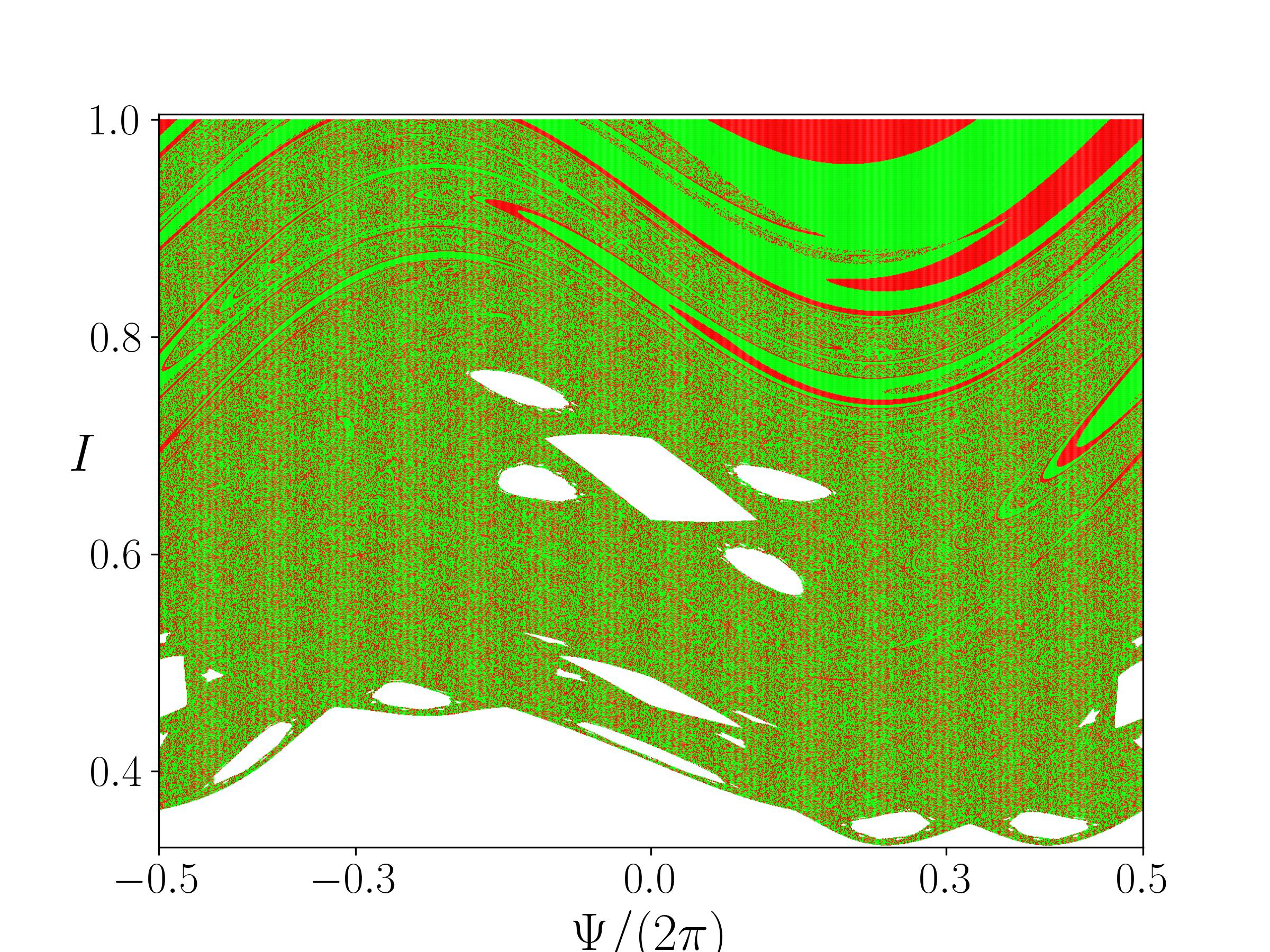}}
    \caption{Escape basins for the exits at $L: (I=1.0, -0.5\leq\Psi/(2\pi) <0.0)$ (green pixels) and $R: (I=1.0, 0.0\leq\Psi/(2\pi)\leq 0.5)$ (red pixels). Those points that do not escape within a maximum allowed time, $n^*=10^5$, are represented by white pixels. The amplitude $\phi$ of the electrostatic fluctuations is (a) $4.92\times 10^{-3}$, (b) $7.65\times 10^{-3}$, (c) $8.74\times 10^{-3}$, and (d) $10.38\times 10^{-3}$. The green and red lines in $I=1.0$ represent the exits $L$ and $R$, respectively.}
    \label{fig:basins}
\end{figure*}

In Figure \ref{fig:basins}(a)-(d), we show the basins of escape of a region of the Poincaré surface of section, for different values of the perturbation strength $\phi$. Points belonging to the basin $L$ are painted green, whereas points of basin $R$ are depicted in red. The white region mostly indicates initial conditions that do not escape within a pre-specified large time $n^*$ (in this case $n^*=10^5$). 

There are regions of white points in the plasma core, which correspond to initial conditions that do not escape (after a maximum time $n^*$), because their trajectories in the phase plane remain on invariant curves, outside the chaotic region. Other white points are inside islands, thus do not escape either. 

The mixing of the escape basins ${\cal B}(L)$ and ${\cal B}(R)$  is clearly seen at most points in the chaotic region. Moreover, the green escape basin, ${\cal B}(L)$, is significantly larger than the red escape basin, ${\cal B}(R)$, for all values we considered for the perturbation amplitude strength $\phi$, indicating a preferential escape through the $L$-exit. This asymmetric feature can be understood by considering the fractal structures that underlie chaotic dynamics in this region, as we describe later on.

\begin{figure*}
    \centering
    \subfloat(a){\includegraphics[height=2.5in]{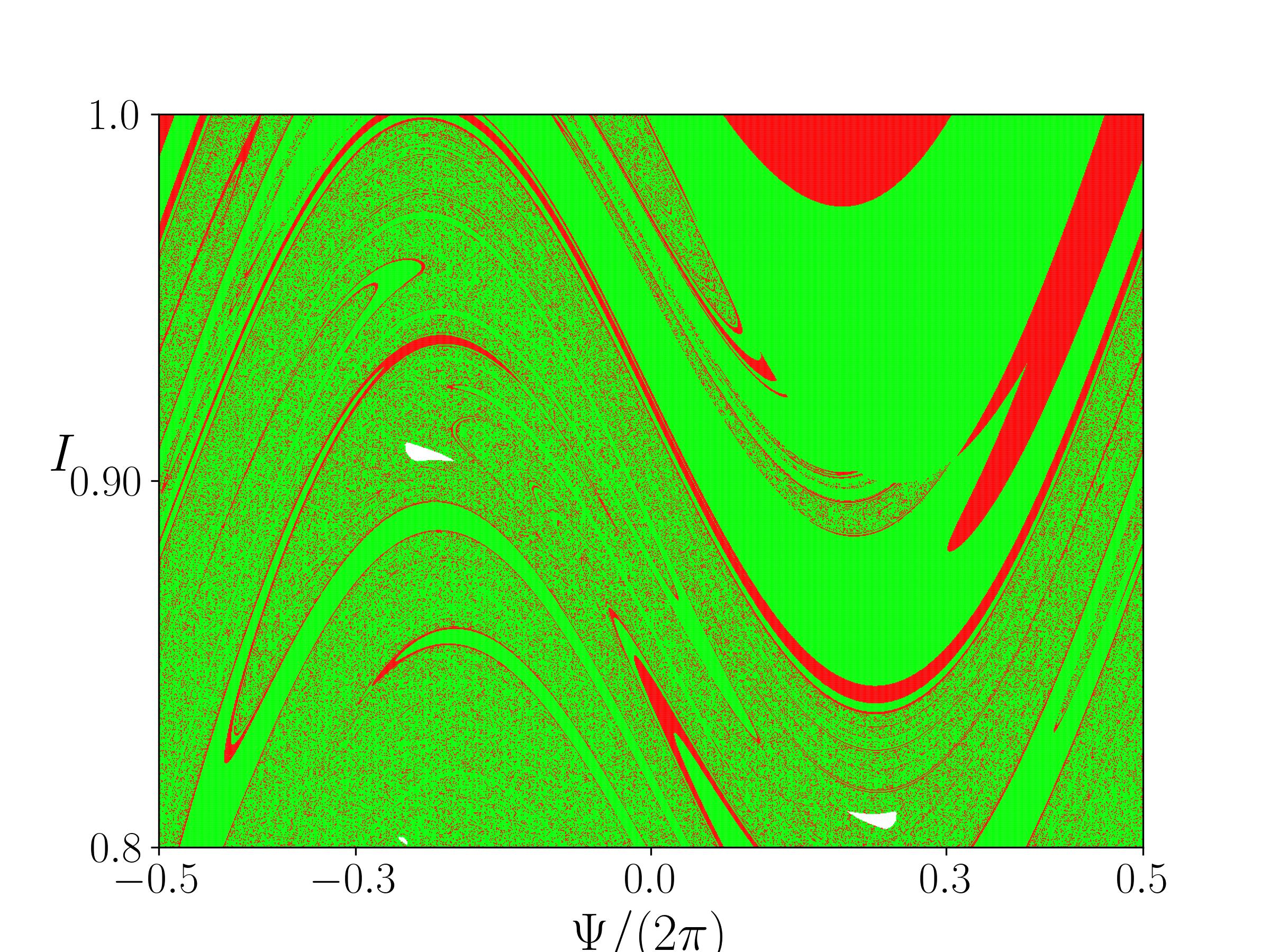}}
    \subfloat(b){\includegraphics[height=2.5in]{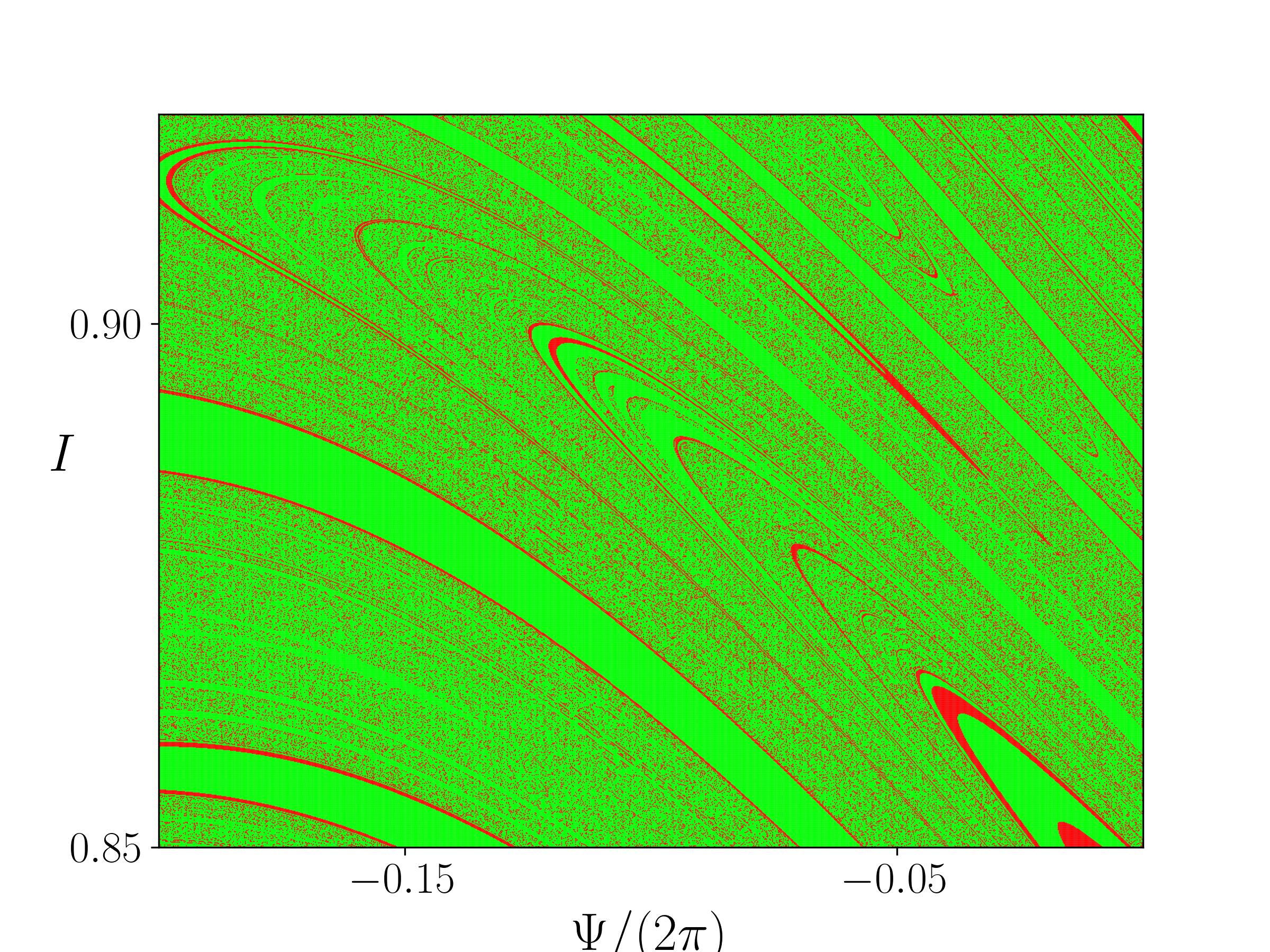}}
    \caption{Two consecutive magnifications of a region of the escape basins obtained for $\phi=8.74\times 10^{-3}$. In (a) is the rectangular region of Fig. \ref{fig:basins}(c), and in (b) we have a further magnification.}
    \label{fig:zoom}
\end{figure*}

The mixing between the two escape basins is non-uniform, as can be seen in FIG. \ref{fig:zoom}(a)-(b), where we show two consecutive magnifications of the escape basins depicted in  FIG. \ref{fig:basins}(c): there is a finger-like structure of red basin filaments embedded in the green basin. 

\begin{figure*}
    \centering
    \subfloat(a){\includegraphics[height=2.5in]{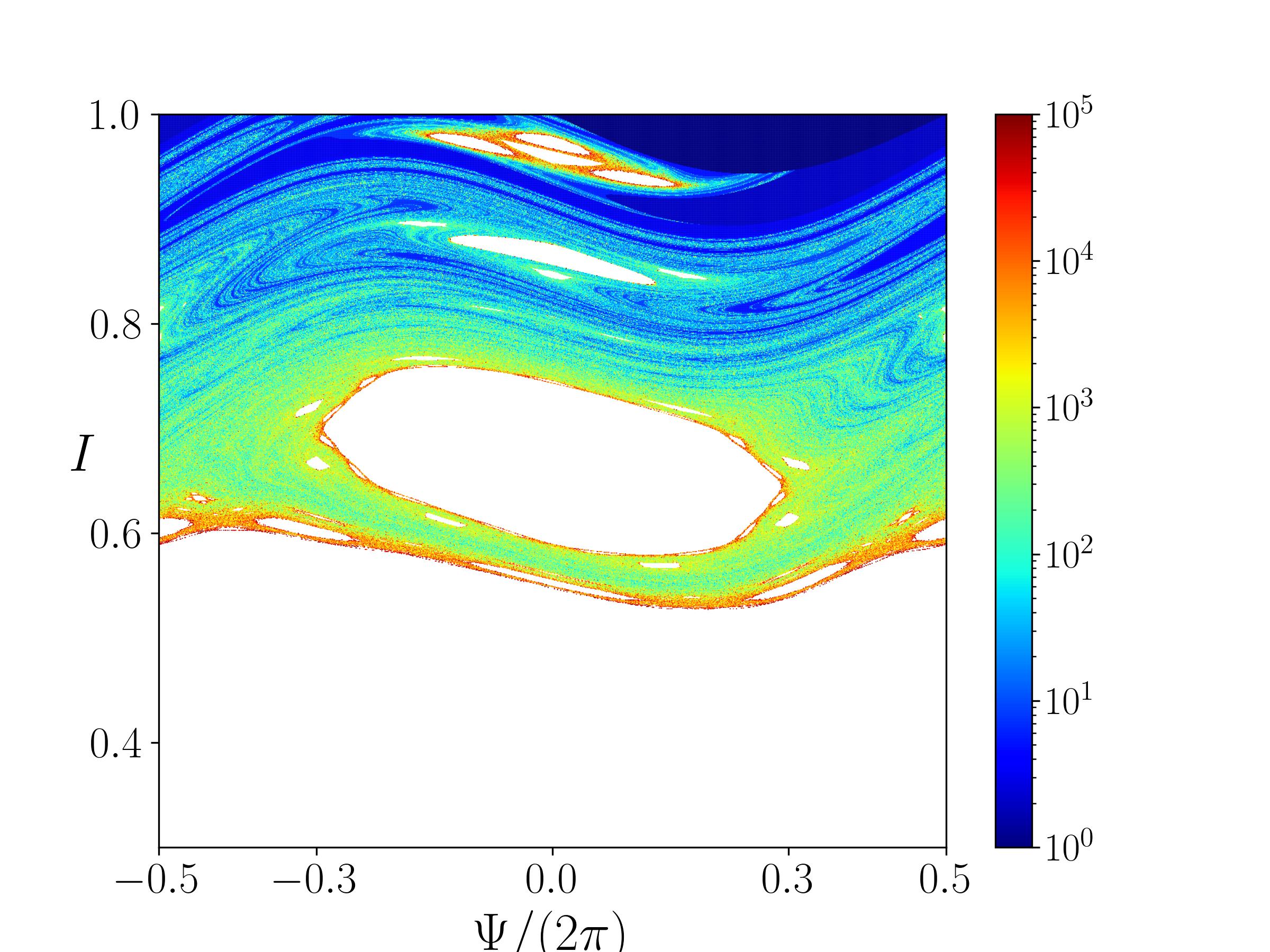}}
    \subfloat(b){\includegraphics[height=2.5in]{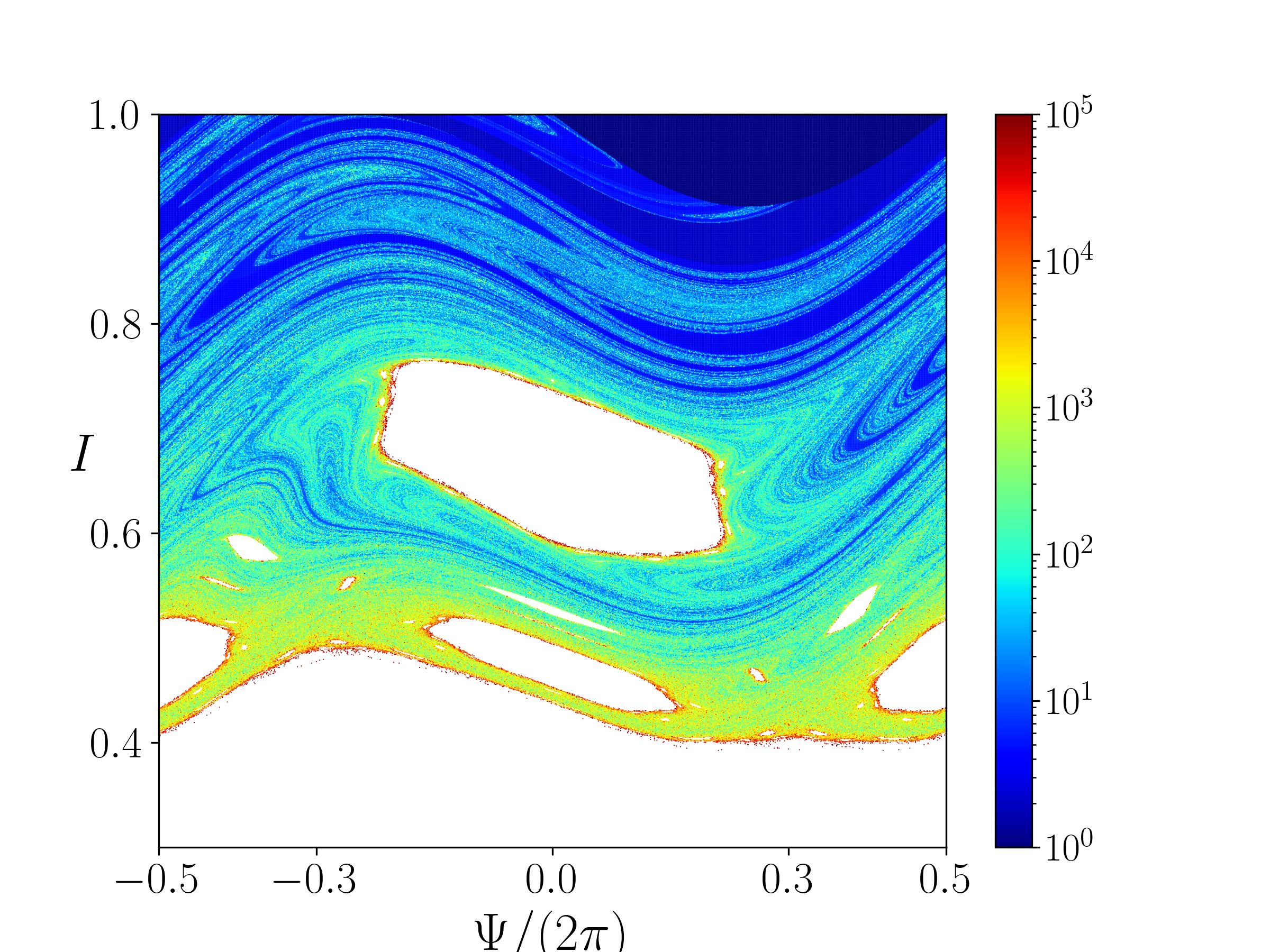}}
    \subfloat(c){\includegraphics[height=2.5in]{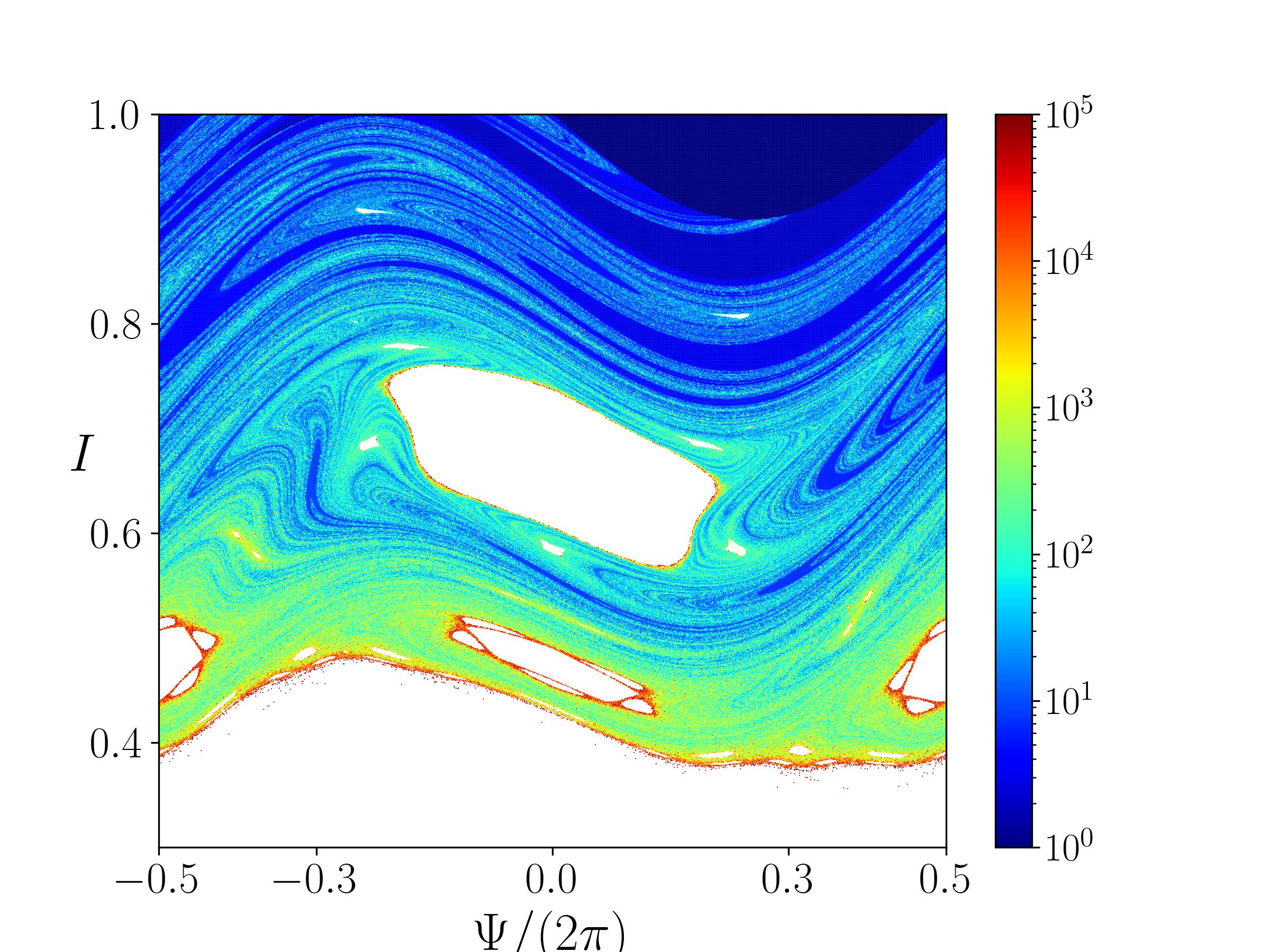}}
    \subfloat(d){\includegraphics[height=2.5in]{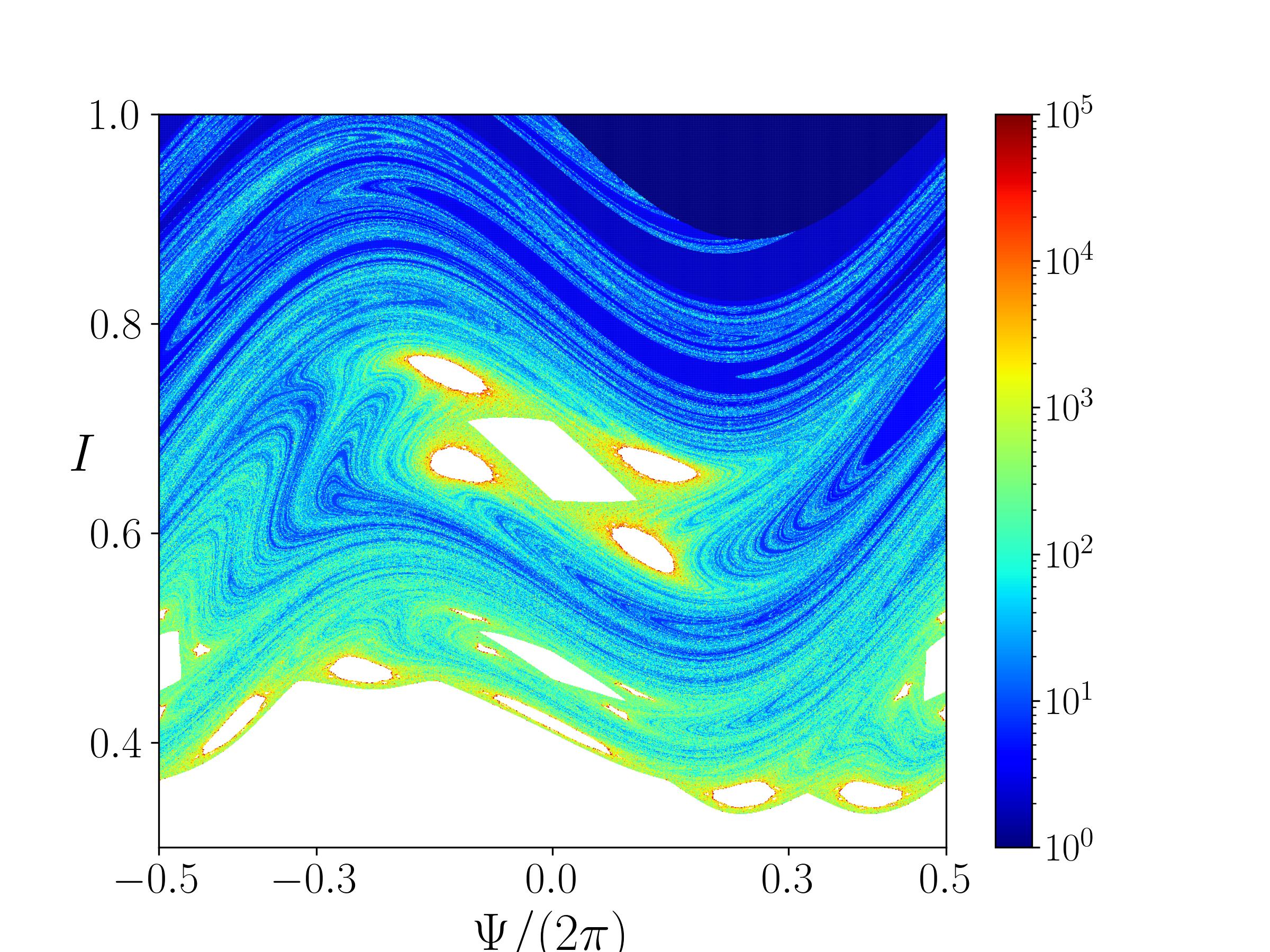}}
    \caption{Escape times (indicated by a colorbar) for different values of the perturbation amplitude $\phi$: (a) $4.92\times 10^{-3}$, (b) $7.65\times 10^{-3}$, (c) $8.74\times 10^{-3}$, and (d) $10.38\times 10^{-3}$.}
    \label{fig:time}
\end{figure*}

Not only the escape basins are intertwined at arbitrarily fine scales, but also the escape time $n_e$, i.e. the number of map iterations that an orbit takes to hit one of the exits, has a complicated distribution in the phase space. Fig. \ref{fig:time}(a), for example, depicts the escape time (in a color bar) as a function of the initial condition $(I,\Psi/(2\pi))$ for the same parameters as the escape basins shown in Fig. \ref{fig:basins}(a). In the chaotic region, the escape time is found to be as finely intermixed as the escape basins themselves. The white points, as before, correspond to points for which the escape time exceeds a specified maximum time $n^*$.  

The chaotic saddle underlying the escaping orbits can be used to understand the complicated structure of escape basins. Let us consider an  unstable periodic orbit embedded in the chaotic region of any phase space shown in Figure \ref{fig:phase}. The stable (unstable) manifold at this point is the set of points whose forward (backward) iterates of the map (\ref{map1})-(\ref{map2}) asymptotically approach each other. The intersections of the stable and unstable manifolds form a non-attracting invariant chaotic set called chaotic saddle \cite{saddle}.

If an initial condition $(I_0,\Psi_0)$ could be placed exactly on an invariant manifold, it would remain on this manifold for arbitrarily large time. However, if this point is off but very close to a given invariant manifold, it would remain so for some time until escape through some of the exits. This property can be used to generate numerical approximations of the invariant manifolds using the so-called sprinkler algorithm \cite{kantz}. Other algorithms for obtaining invariant manifolds are available, but this particular one is easier to apply since one does not need to consider inverse images of the points \cite{invariants}.

Let us consider a bounded region $\mathcal{R}$ of the phase space $I \times \Psi$ containing a chaotic orbit, and cover it with a fine grid of points. Each mesh point corresponds to an initial condition $(I_0,\Psi_0)$, and it is iterated $m$ times using the map (\ref{map1})-(\ref{map2}). After $m$ iterates, if the value of $(I_m,\Psi_m)$ remains inside $\mathcal{R}$, the corresponding initial conditions are numerical approximations of a branch of the stable manifold $W^s(P)$ which emanates from an unstable periodic orbit $P$ embedded in the chaotic orbit in the region $\mathcal{R}$. Moreover, the $m$-th iterates themselves, $(I_m,\Psi_m)$, are numerical approximations of the unstable manifold $W^u(P)$. Analogously, the corresponding $m/2$-th iterate constitutes a numerical approximation of the chaotic saddle itself \cite{poon}. The underlying chaotic behavior of the system can be explained by the chaotic saddle, whose topological properties are similar to the Smale horseshoe \cite{smale}. 

\begin{figure*}
    \centering
    \subfloat(a){\includegraphics[height=2.3in]{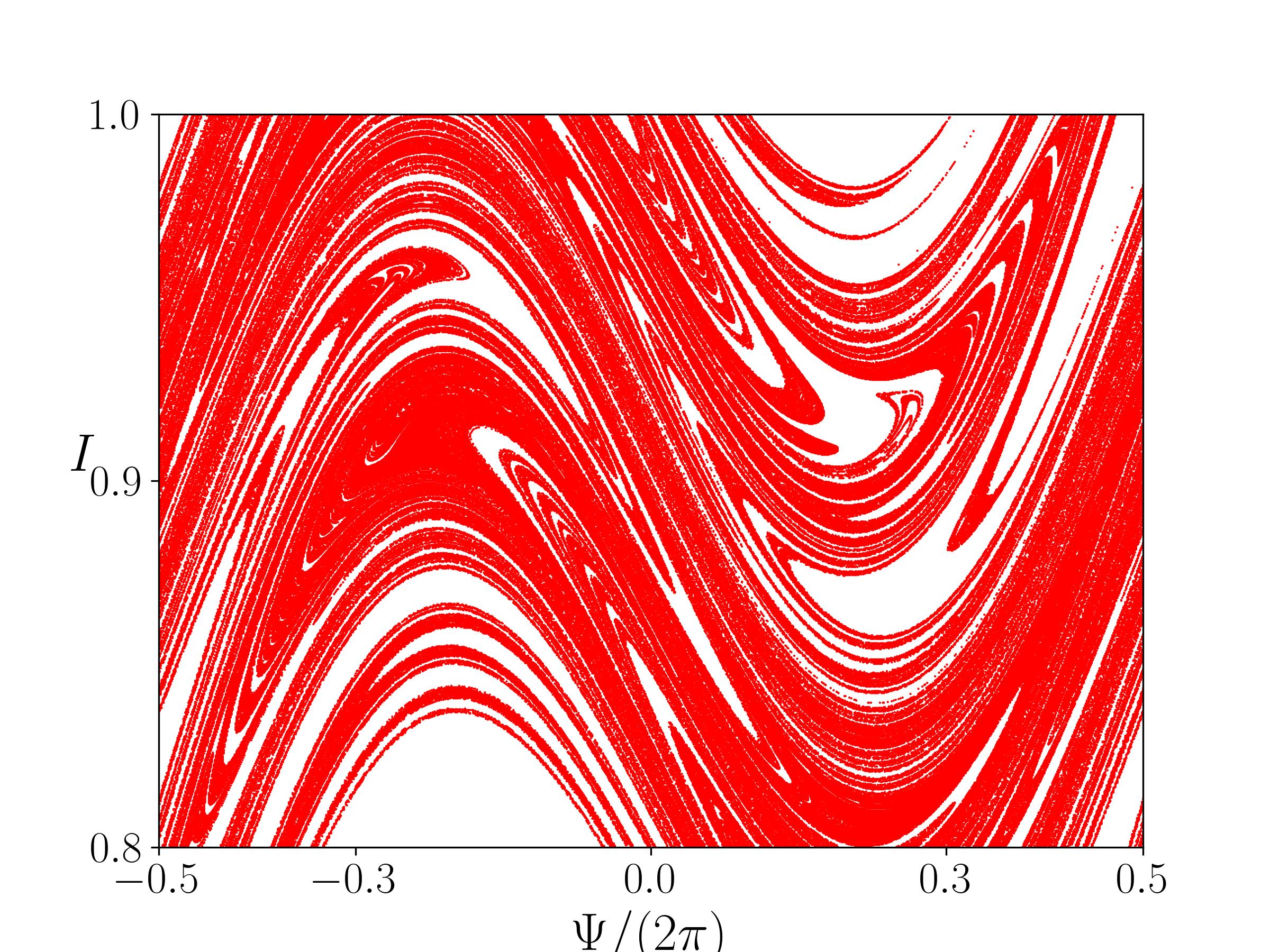}}
    \subfloat(b){\includegraphics[height=2.3in]{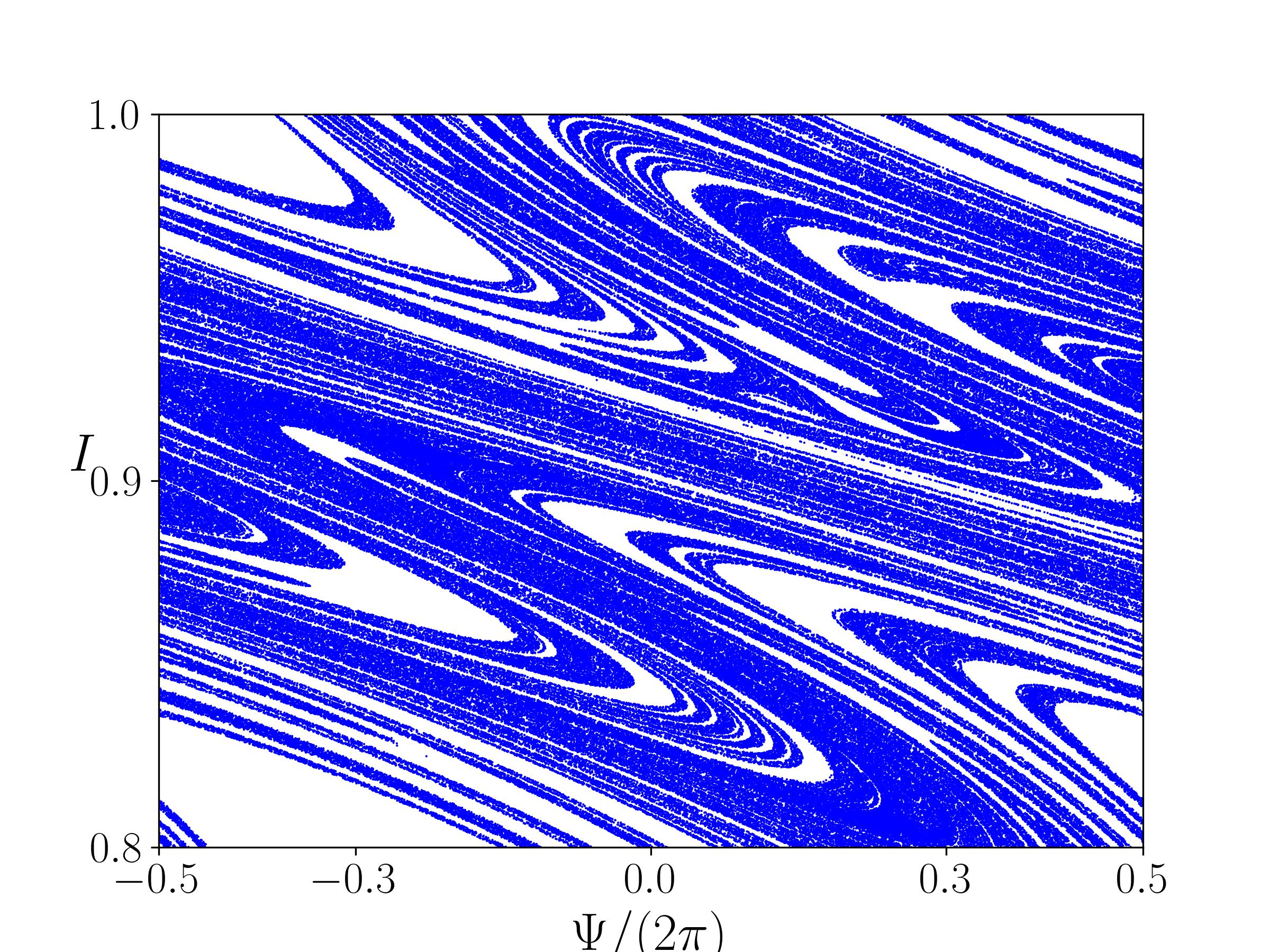}}
    \subfloat(c){\includegraphics[height=2.3in]{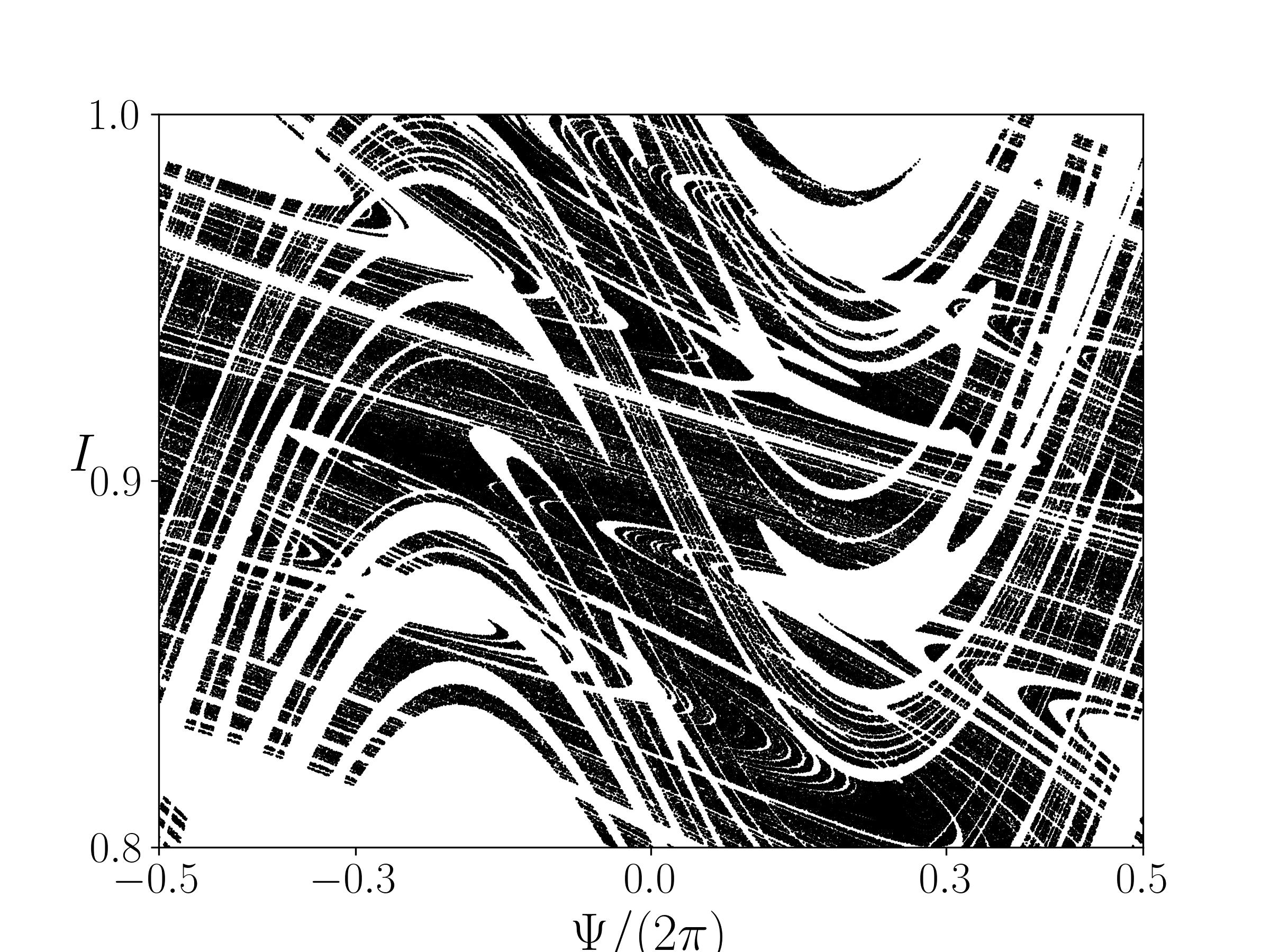}}
    \caption{Numerical approximations of the (a) stable and (b) unstable invariant manifolds for an unstable fixed point embedded in the chaotic region when $\phi=8.74\times 10^{-3}$. The points in (c) are numerical approximations of  the corresponding chaotic saddle. The calculation was done with a grid of $1000\times 1000$ initial conditions with $m=10$ iterations.}
    \label{fig:manifolds}
\end{figure*}

In FIG. \ref{fig:manifolds} we show numerical approximations of stable and unstable manifolds, for $\phi = 8.74\times 10^{-3}$, obtained by the sprinkler method. We used a grid of $1000\times 1000$ initial conditions with $m=10$ iterations for our simulations. The boundary of the escape basins coincides with the stable manifold of the chaotic saddle as can be seen in Figures \ref{fig:zoom}(a) and \ref{fig:manifolds}(a), as explained by Lai and Tél\cite{lai-tel}.

The unstable manifold shown in FIG. \ref{fig:manifolds}(b) indicates the path followed by map iterations before they escape, i.e. the escape channels by which particles pass toward the tokamak wall. If an initial condition $(I_0,\Psi_0)$ is off but arbitrarily near the unstable manifold, it will remain near it for an arbitrarily long time. 


\section{Characterization of fractal structures}

\subsubsection{Uncertainty exponent}

In order to characterize the fractality of escape basins, we first calculated the uncertainty dimension according to the algorithm introduced by MacDonald {\it et al.} \cite{macdonald,macdonald1}. Any initial condition in the phase space is known with some uncertainty that we can represent by a disk of radius $\varepsilon$ centered at  $(I_0, \Psi_0)$. If the disk intercepts the boundary of the escape basins, we call that initial condition $\varepsilon$-uncertain, i.e. it is impossible to predict with total confidence by which exit that initial condition will escape through, if it is specified with uncertainty $\varepsilon$. This impossibility is called final state uncertainty and it is directly related to the fractal nature of the escape basin boundary. 

We consider a grid of initial conditions $(I_0, \Psi_0)$ in a given phase space region ${\cal R}$ containing a significant portion of the escape basin boundary. The points belonging to this grid are taken to be the centers of small disks of radius $\varepsilon$, and are iterated until the ensuing orbit escapes through $L$ or $R$ exits (if the orbit does not escape at all, it is discarded from the computation). For each grid point, two other initial conditions are randomly chosen inside the corresponding  $\varepsilon-$disk and they are iterated again until reaching one of the two exits. If one of the three points, for a given $\varepsilon-$disk, fails to escape through the same exit, the center of this disk is considered $\varepsilon-$uncertain. 

The uncertain fraction $f(\varepsilon)$ is the number of $\varepsilon-$uncertain conditions divided by the total number of initial conditions. It is known that $f$ scales with $\varepsilon$ as a power law $f(\varepsilon)\sim\varepsilon^\xi$, where $\xi$ is called the uncertainty exponent. Let $d$ be the box-counting dimension of the escape basin boundary in the two-dimensional phase plane. In order to cover the boundary with boxes of length $\delta$, it takes $N(\delta) \sim \delta^{-d}$ of them, so that the box-counting dimension is given by 
\begin{equation}
    \label{boxcount}
    d = \lim_{\delta\rightarrow 0} \frac{\ln N(\delta)}{\ln (1/\delta)}.
\end{equation}

Now we set $\delta$ equal to the initial condition uncertainty $\varepsilon$, and thus the area of the uncertain region of the phase space will be of the order of the total area of all $N(\delta)$ boxes used to cover the escape basin boundary. Given that the area of each box is $\varepsilon^2$, the uncertain area is of the order 
\[
f(\varepsilon) \sim \varepsilon^2 N(\varepsilon) = \varepsilon^{2-d} = \varepsilon^\xi,
\]
so that the escape basin boundary dimension is $d = 2 - \xi$.  If the escape basin boundary is a smooth curve ($d = 1$), then $\xi=1$. However, if the basin boundary is fractal, then $0<\xi<1$, so that its dimension is $1<d<2$.

In our simulations, we used a grid of $10^4\times10^4$ initial conditions placed in the chaotic region of the phase plane $I\times\Psi$ and iterated $10^5$ times. If the initial condition does not escape after this number of iterations, it is removed from the computation. For each value of $\varepsilon$, we repeat ten times the computation of the uncertainty fraction, the local error being the standard deviation of the results. Ten values of $\varepsilon$, namely $10^{-k}$ ($1 \leq k \leq 10$) are used to make a diagram of $\log{f(\varepsilon)}$ versus $\varepsilon$, the uncertainty dimension is determined by least-square fits. The global error is the average local error for each $\varepsilon$. Our results, for different values of $\phi$, are summarized in Table \ref{tab:dimension}. The uncertainty dimension varies very little with $\phi$ and is very close to $2.0$, which is the limiting case of an area-filling curve. In all those cases, the basin boundary is extremely mixed. These results point to an extreme fractal escape basin structure. 

\begin{table}
    \centering
    \begin{tabular}{|c|c|c|c|} \hline 
         $\phi\,\,\,(10^{-3})$ &  $\xi$ & $d$ & error \\\hline\hline 
         4.92   & 0.001     & 1.999   &   0.001  \\\hline
         7.65   & 0.001    & 1.999   &   0.001  \\\hline
         8.74   & 0.003    & 1.997   &   0.002     \\\hline
         10.38  & 0.020    & 1.980   &   0.030  \\\hline  
    \end{tabular}
    \caption{Uncertainty exponents and box-counting dimensions for the escape basin boundary for different values of the perturbation amplitude.}
    \label{tab:dimension}
\end{table}

\subsubsection{Basin entropy}

We used the concept of basin entropy \cite{daza-entropy} to quantify the final state uncertainty produced by the fractality, using ideas of information theory. We considered a bounded region $\mathcal{R}$ in the chaotic region of the phase space in Fig. \ref{fig:phase}, characterized by the presence of $N_A$ exits. We divided $\mathcal{R}$ in a fine mesh of $N$ boxes, each of which containing a grid of $\zeta \times \zeta$ sample initial conditions. The map associates to each initial condition on the grid a single variable (called a color) labeled from 1 to $N_A$. The basin entropy can be obtained from computing the information entropy for the boxes.  

The color in each grid point represents the value of an integer (pseudo-)random variable $j$. Let $p_{ij}$ denote the probability that the $j$th color is assigned  to the $i$th box, i.e. the frequency of color $j$ among the $\zeta^2$ initial conditions in box $i$. Treating the chaotic orbits of our map as statistically independent, the basin entropy of the $i$th box is defined as
\begin{equation}
\label{eq:si}
    S_i = -\sum_{j=1}^{N_A} p_{ij} \, \log{p_{ij}},
\end{equation}
 with $0 \log 0 = 0$ by convention. The total basin entropy for the region $\mathcal{R}$ is then
\begin{equation}
\label{eq:Sb}
    S_b = \frac{1}{N}\sum_{i=1}^N S_i.
\end{equation}

In the case of only one exit, $S_b=0$ and there is no uncertainty in the final state caused by fractality. Moreover, if there are $N_A$ equiprobable exits, the basin entropy assumes the maximum value $S_b = \log{N_A}$, completely characterizing the escape basin structure. We also adapt this entropy calculation to evaluate the uncertainty related to the escape basin boundary. In order to do this, we repeat the same calculation described above, but considering only the $N_b$ boxes that contain more that one color, i.e. if a box $i$ contains only one color, we disregard $i$ in the calculation of the entropy.  In this way, noting also that $S_i = 0$ for single-color boxes, we compute the basin boundary entropy as $S_{bb} = (1/N_b) \sum_i S_i = N \, S_b / N_b$.

In our case, there are two exits $L$ and $R$, and the region $\mathcal{R}$ is the rectangle $0.3\leq I\leq 1.0$, $-0.5 \leq \Psi/(2\pi) \leq0.5$ covered with a grid of $1000\times 1000$ initial conditions, distributed inside $4\times 10^4$ boxes, $\zeta=5$. For each box, we computed a maximum of $10^5$ iterations of the map for a number of initial conditions therein, the orbits that do not escape up to this time being excluded from the statistics. Let $n_L$ and $n_R$ denote the number of points in each grid cell that escape to exits $L$ and $R$, respectively. The probability for the $i$th box is 
\begin{equation}
    p_{iL} = \frac{n_L}{n_L + n_R}, \qquad  
    p_{iR} = \frac{n_R}{n_L + n_R}, 
\end{equation}
so that the entropy for that grid cell is $S_i = -p_L\log{p_L} - p_R\log{p_R}$. Summing up over the entropy of all boxes and dividing by the number of boxes, we obtain the basin entropy $S_b$. The basin boundary entropy is obtained, excluding from the summation those boxes for which either $p_L=0$ or $p_R=0$. Since there are two exits, $S_b$ and $S_{bb}$ vary between $0$ and $\log{2}\approx 0.69$. 

\begin{figure}
    \centering
    \includegraphics[height=2.5in]{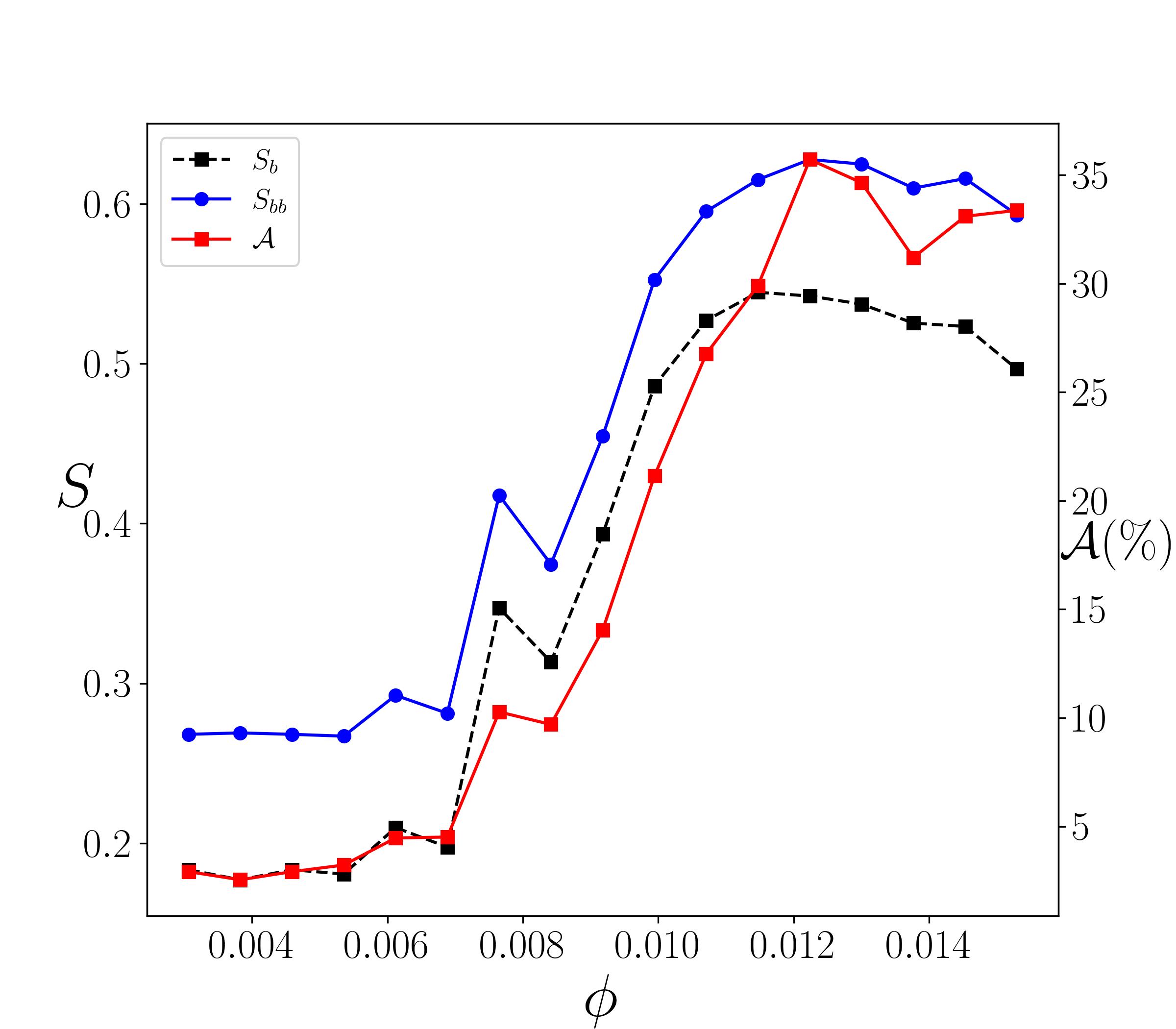}
    \caption{Escape basin entropy (blue), basin boundary entropy  (black), and relative area ${\cal A}$ of the red escape basin ${\cal B}(R)$ (red) as a function of the perturbation strength $\phi$.}
    \label{fig:entropy}
\end{figure}

In Fig. \ref{fig:entropy} we show our results for the basin and boundary entropies as a function of the amplitude of the drift waves $\phi$. As a trend, the entropies increase with $\phi$, this means that the red and green basins become progressively more mixed and involved. Moreover, we see that both $S_b$ and $S_{bb}$ follow the increase of the occupied area of the red basin as depicted in Figure \ref{fig:entropy}. Inspecting Figures \ref{fig:basins}, we see that the green is predominant over the red basin, but with the increase of $\phi$ the red area becomes larger. The entropies increase as the red basin expands and becomes comparable to the green basin. 

\section{Wada property}

The fractal structures we have discussed so far are related to boundaries between two different exit basins. However, it is interesting to investigate the case of three or more basins. Extending the previous reasoning, one would expect to find fractal structures in the corresponding boundaries, but for three or more basins an important question to be answered is: are almost all boundaries separating points of just two basins? If the answer is negative, conceptual problems may appear since the basins are restricted to a limited phase space domain. As we will see, not all boundary points separate just two basins, and there is a considerable number of boundary points that separate simultaneously three or more basins, which is a non-trivial topological concept called Wada property\cite{yorke-1991}. 

In order to discuss this result, some preliminary definitions are needed. Let the system have more than one escape basin. If a given point has a neighborhood consisting just of points belonging to a single escape basin, it is called an interior point. A point $P$ is a boundary point of the basin $\mathcal{B}$ if every open neighborhood of $P$ intersects both basin $\mathcal{B}$ and at least another basin $\mathcal{B'} \ne \mathcal{B}$. If the point $P$ is a boundary point of at least three different basins, then we say that $P$ is a Wada point. If the escape basin boundary is a fractal curve, then a fraction of its points can be Wada points, so that we say the boundary has the Wada property (partially or totally). 

Boundaries possessing the Wada property have important physical consequences, given that a boundary point turns to be arbitrarily close to points of at least three basins of escape \cite{yorke-1991}. Since an initial condition is always known up to a given uncertainty, in a system with the Wada property, it is not possible to say with certainty to which exit a particle will escape. Hence, the Wada property is an extreme form of final-state sensitivity.

\begin{figure*}
    \centering
    \subfloat(a){\includegraphics[height=2.5in]{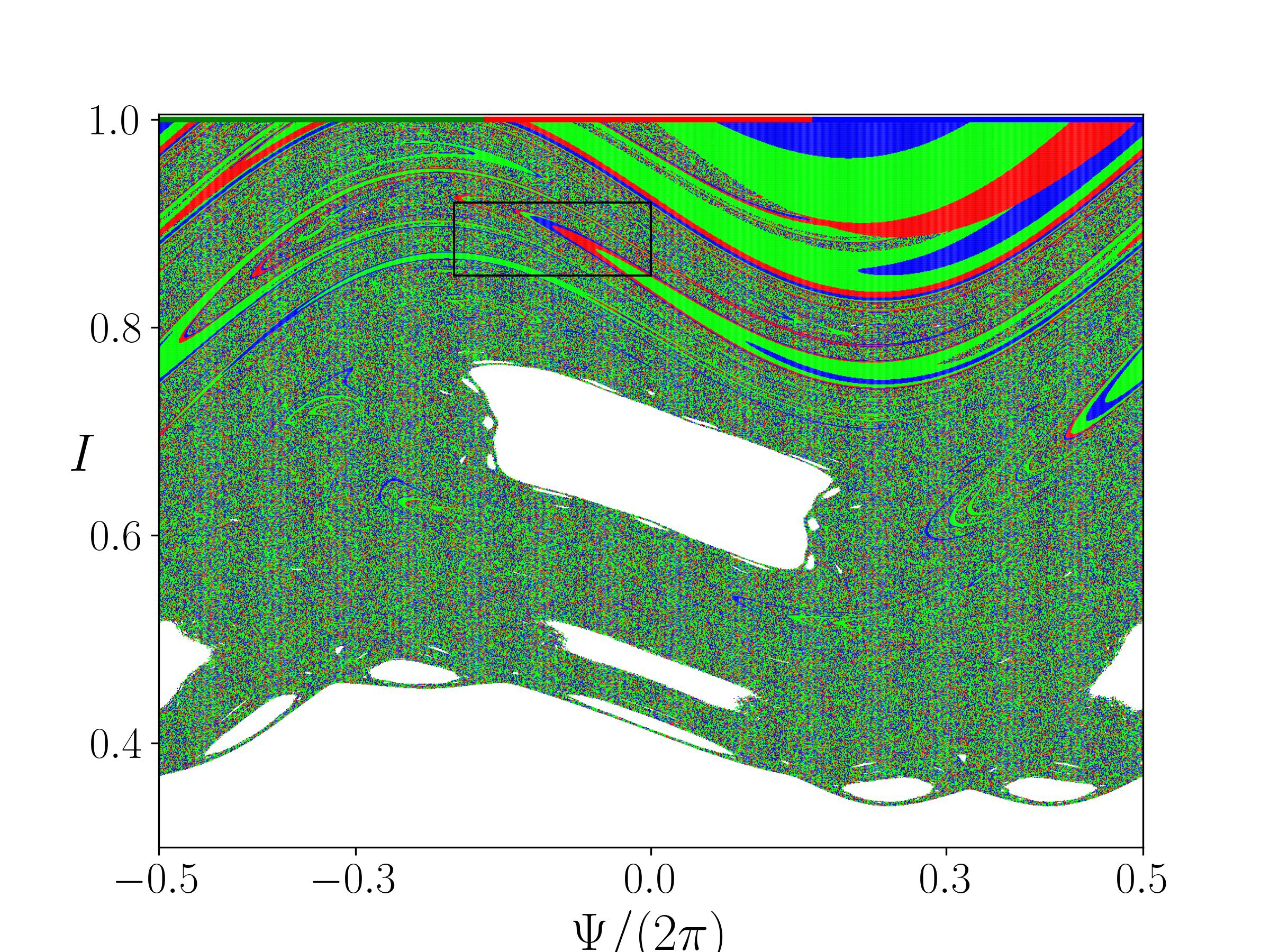}}
    \subfloat(b){\includegraphics[height=2.5in]{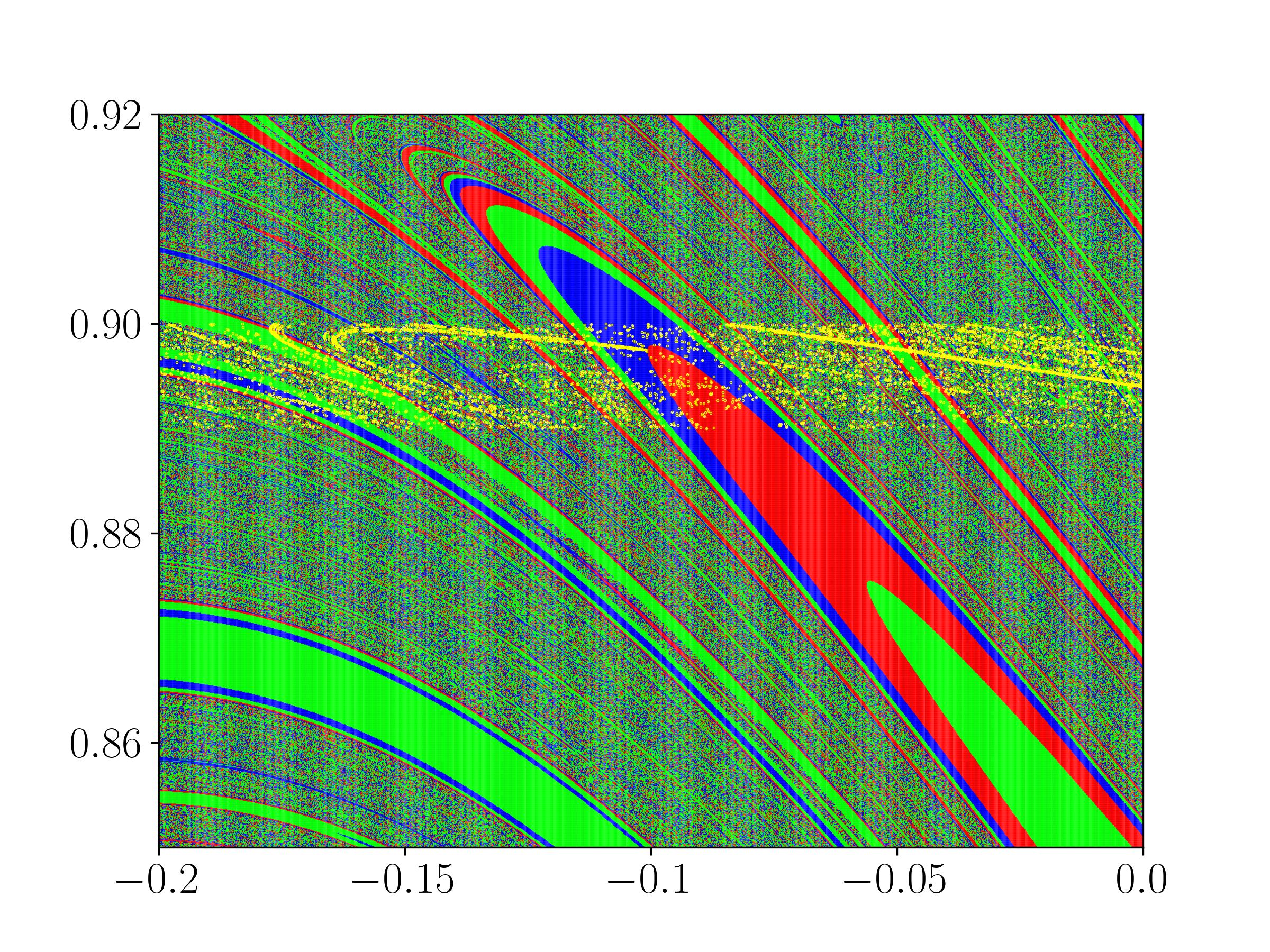}}
    \caption{(a) Basins of escape for the case of three exits. Points belonging to the basins of exits $L$, $C$, and $R$ are painted red, blue, and green, respectively. The green, blue and red lines in $I=1.0$ represent the exits $L$, $C$ and $R$, respectively. (b) A magnification of the black rectangle drawn in (a). The  yellow points are a numerical approximation of the invariant unstable manifold crossing all escape basins.}
    \label{fig:wada}
\end{figure*}

We considered three exits by dividing the tokamak wall $I = 1.0$ into three congruent segments denoted by $L: -\pi < \Psi \leq -\pi/3$, $C: -\pi/3 < \Psi \leq \pi/3$, and $R: \pi/3 < \Psi \leq \pi$. We evaluated the corresponding escape basin for these exits using the same procedures already described for two exits.  Our results are shown in FIG.  \ref{fig:wada}(a), for a perturbation amplitude $\phi= 9.84\times 10^{-3}$. Points belonging to the basins of $L$, $C$, and $R$ are painted red, blue, and green, respectively. The corresponding escape basins have a similar shape as those described before, with a fingerlike structure, and they also seem to be densely intertwined, but it is difficult (if not impossible) to discern the Wada property just by a cursory inspection. 

In order to test the Wada property, in the escape basins produced by the map (\ref{map1})-(\ref{map2}), we have to prove that the unstable manifold of a periodic orbit intersects all the escape basins. This is a necessary but not sufficient condition to have the Wada property fulfilled \cite{nusse-yorke-1996}. Since the rigorous demonstration of this property is not feasible for the map we are dealing with, we rely on numerical signatures of such behavior. In FIG. \ref{fig:wada}(b) we indicate that the unstable manifold emanating from an unstable fixed point, embedded in the chaotic region, intersects the three escape basins, what strongly suggests that the escape basins fulfill the Wada property. However, such evidence does not inform what is the fraction of boundary points that are Wada points. 

In order to characterize which boundary points have the Wada property, we used the so-called grid approach \cite{grid-approach}. Let ${\cal R}$ be a bounded region of the phase space (mostly in the chaotic region near the tokamak wall) containing $N_A\geq 3$ exits, and let us denote by  $\mathcal{B}_j$, $j=1,2,\dots N_A$, the corresponding basins of escape. Using a fine rectangular mesh, the region ${\cal R}$ is divided into a set of non-overlapping boxes ${b_1, b_2,\dots, b_k}$. We iterate each point $(x,y)$ of ${\cal R}$ in order to find which exit the particle escapes through, so as to determine the corresponding escape basin $\mathcal{B}_j$.

We defined $C(x,y) = j$ if $j\in\mathcal{B}_j$ and $C(x,y) = 0$ if $(x, y)$ is in none of the sets, the value of $C$ is the color of the box. For any rectangular box, we define $C(b) = C(x,y)$ where $x,y$ is the point at the center of box. We define $P(b_j)$ the collection of grid boxes consisting of $b_j$ and all boxes that have at least one point in common with $b_j$, in the two dimensional case $P(b_j)$ is a $3\times 3$ collections of boxes with $b_j$ as the central box. The number of different $C(b_j)$ (colors) in $P(b_j)$ for each $b_j$ is $M(b_j)$. Provided $M(b_j)\neq 1, N_A$ for a given $b_j$, we take the two closest boxes in $P(b_j)$ with different colors and draw a line segment between them, calculating the color of the midpoint of this line. If the color of the segment midpoint is such that we have all the possible colors inside $P(b_j)$, then $M(b_j)=N_A$ and we stop the procedure. Otherwise, we choose intermediate points in this line segment and repeat this procedure until $M(b_j) = N_A$, unless the number of points exceeds a specified limit.  

After having obtained the values of $M(b_j)$ for all grid points, we determine the set $G_m$ of those original boxes such that $M(b_j) = m$, for a given integer $m$. If $m=1$, we have the set $G_1$ containing points belonging to the interior of the $j$th escape basin (interior points). Analogously, the set $G_2$ contains points belonging to the boundary between two escape basins, that is, there are two different colors inside the set  $C(b_j)$ (boundary points). In the same way, $G_3$ consists of points that belong to the boundary between three basins, i.e. the set $G_3$ contains Wada points (points satisfying the Wada property).

Since the procedure outlined involves a number of refinements, let us denote by $G_m^q$ the set $G_m$ obtained at the $q$-th  procedure step. We expect that, as $q$ goes to infinity, the sequence of refinements converge to a final set $G_m$, in such a way that we compute the following quantity
\begin{equation}\label{eq:W_m}
    W_m = \lim_{q\rightarrow\infty} \frac{\mathcal{N}(G_m^q)}{\sum_{j=2}^{N_A}\mathcal{N}(G_j^q)}, \qquad (m = 2, 3, \ldots N_A),
\end{equation}
where $\mathcal{N}(G_j^q)$ is the number of points of the set $G_j$ at the $q$-th refinement step. 

In the case of $W_m=0$, the system has (almost) no grid boxes that belong to the boundary separating $m$ escape basins. If $W_m=1$, then (almost) all the boxes belong to the common boundary of $m$ escape basins. The system is said to have the Wada property if $W_{N_A}=1$, given that it is always possible to find any color arbitrarily close to the boundary between two other colors. The system is said to be partially Wada when $0<W_{m}<1$, with $m\geq 3$.

In our problem, with $N_A=3$ escape basins, we calculated $W_2$ and $W_3$ for an increasing number $q$ of procedure steps of computing colors at the intermediate points between adjacent boxes, namely
\begin{align}
    W_2 &= \frac{\mathcal{N}(G_2)}{\mathcal{N}(G_2) + \mathcal{N}(G_3)}, \label{eq:W_2} \\
    W_3 &= \frac{\mathcal{N}(G_3)}{\mathcal{N}(G_2) + \mathcal{N}(G_3)}. \label{eq:W_3}
\end{align}
We checked, for each $q$th iteration of the procedure, whether or not points of $G_2$ may belong to $G_3$ by testing $2(q-2)$ initial conditions which are intermediate between the central box and a neighbor box with different color. If some of these initial conditions present the missing color, the central box is reclassified as $G_3$. 

\begin{figure*}
    \centering
    \subfloat(a){\includegraphics[height=2.5in]{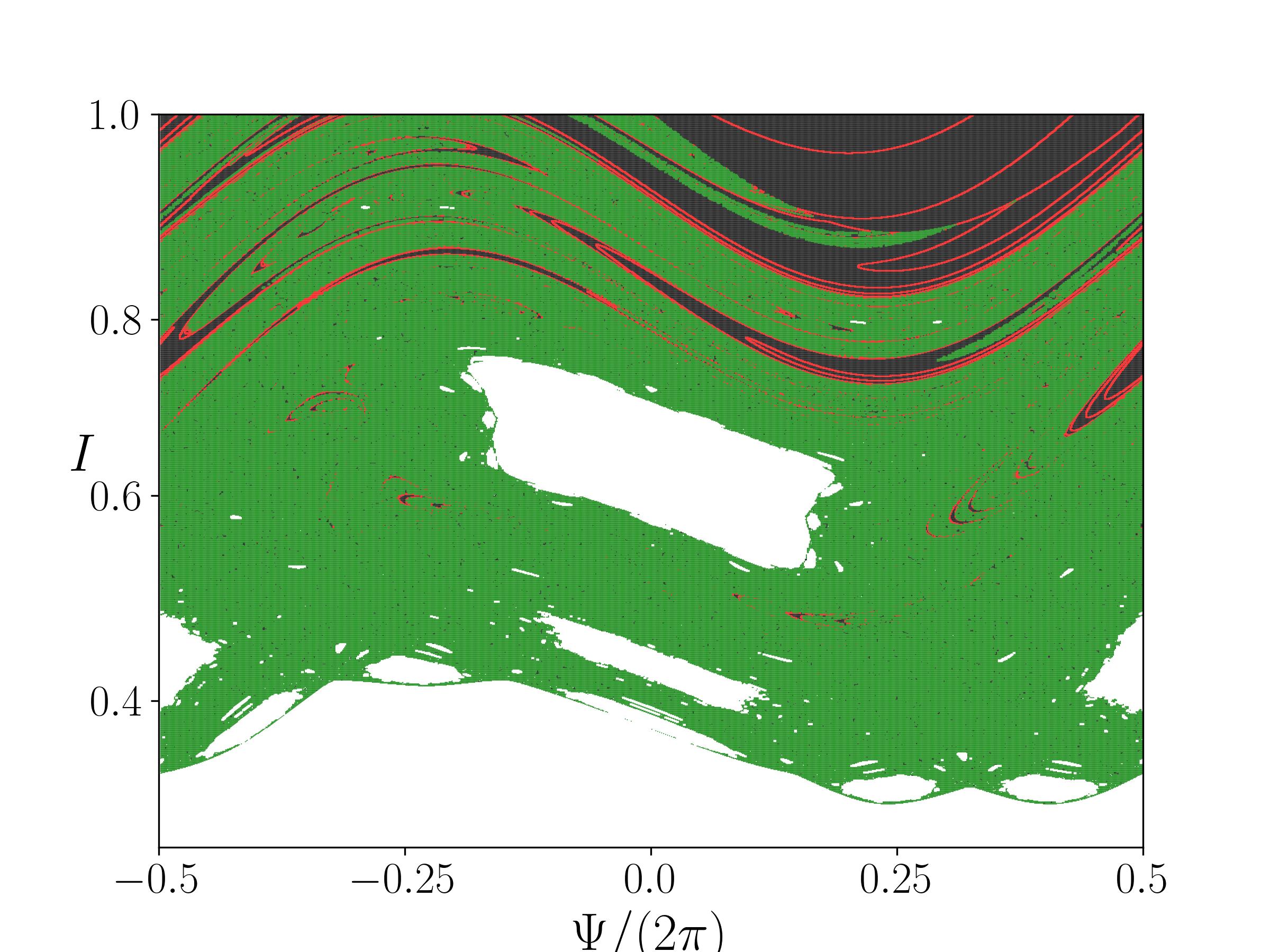}}
    \subfloat(b){\includegraphics[height=2.5in]{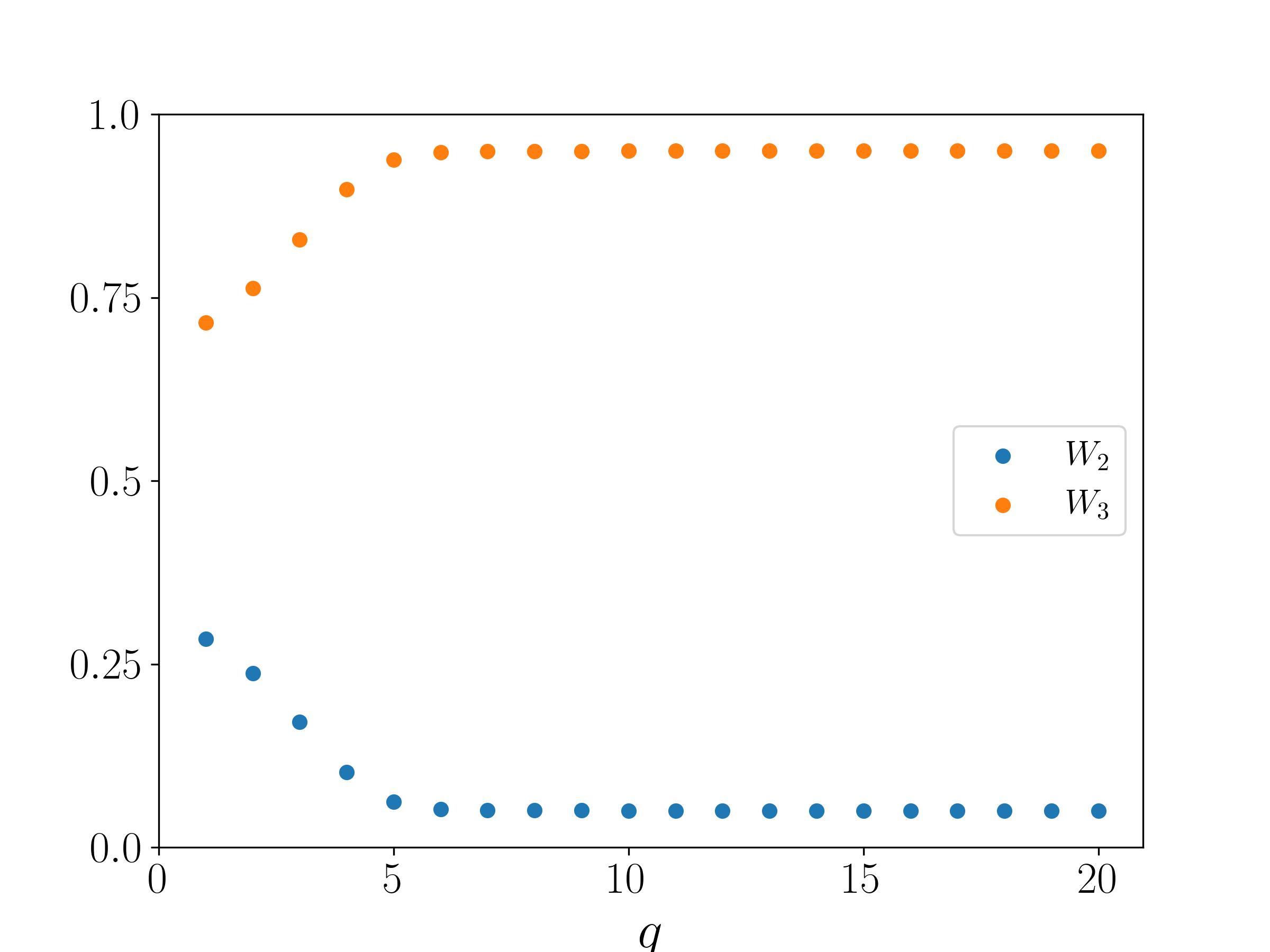}}
    \subfloat(c){\includegraphics[height=2.5in]{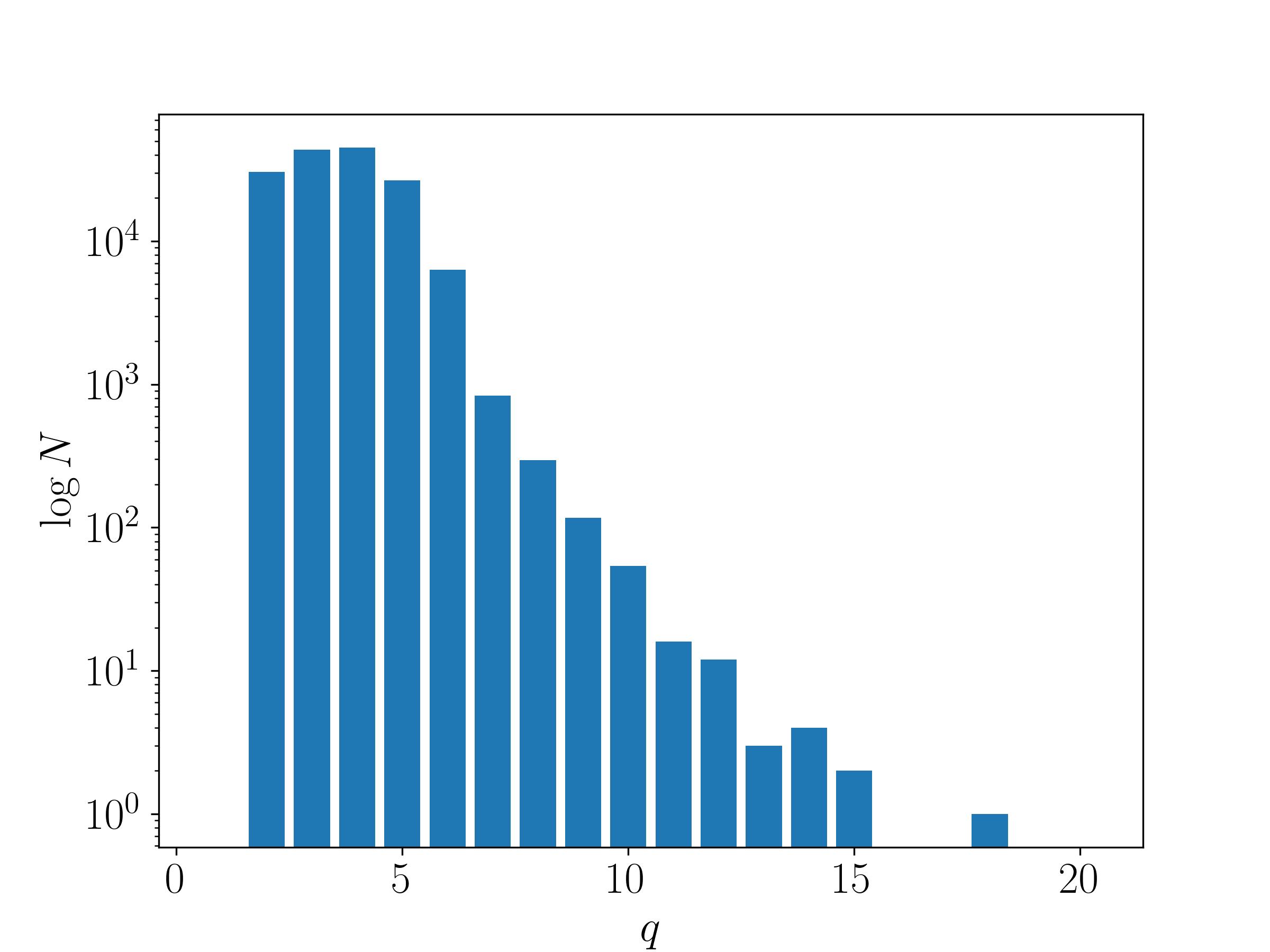}}
    \caption{(a) Basin structure of Figure \ref{fig:wada}(a), showing points belonging to the $G_1$ set (interior points, black), points of the $G_2$ set (boundary points between two basins, red), and points of the $G_3$ set (boundary points between three basins, green), after $q = 20$ refinement steps. (b) Values of the quantities $W_2$ (blue) and $W_3$ (orange) as a function of the refinement step. (c) Histogram (semilog) showing the number of reclassified points for various numbers of refinement steps.}
    \label{fig:grid}
\end{figure*}

Our results, after $20$ refinement steps, are show by Figure \ref{fig:grid}(a), where we plot the points classified as $G_1$ (black points), $G_2$ (red points) and $G_3$ (green points). We observe a predominance of Wada points belonging to the set $G_3$, in agreement with the complex basin structure displayed by Fig. \ref{fig:wada}. Curiously the number of interior points is relatively small, as well as those belonging to a boundary between only two escape basins. This suggests that the Wada property holds in quite a large degree for our system. 

The values of $W_2$ and $W_3$ are shown in Figure \ref{fig:grid}(b) as a function of $q$. We observe a fast convergence after just $q=4$ iterations, yielding $W_2\approx 0.0424$ and $W_3 \approx 0.9576$. Hence the basins of escape are  partially Wada but, since {\it circa} $96 \%$ of the boundary points are Wada points, the system is close to being totally Wada. The fast convergence can be also appreciated in Figure \ref{fig:grid}(c), which shows a histogram for the number of points initially classified as belonging to the set $G_2$ but which are reclassified to the $G_3$ set at each refinement step, after a large number of evaluations of the quantities $W_2$ and $W_3$. We see that most of the convergence is obtained after $3$ to $5$ steps, and the number of reclassified points decreases exponentially to zero as $q$ increases. 

\section{Collisional map}

In the theoretical model that we used to describe the drift motion of impurity particles, we neglected their mutual interaction through collisions and others processes involving the particles interaction with the plasma, an approximation valid if the collisional frequency is low enough. However, in cases where this assumption may be not entirely true, some kind of collisional effect should be taken into account. However, since our description is limited to the dynamics in the Poincaré surface of section, a detailed treatment of collisions is not feasible (using e.g. scattering functions for a given cross-section). We consider instead a phenomenological model in which the essential feature of collisions is implemented as random perturbations of the particle position in the Poincaré surface of section. 

\begin{figure}
    \centering
    \subfloat(a){\includegraphics[height=2.5in]{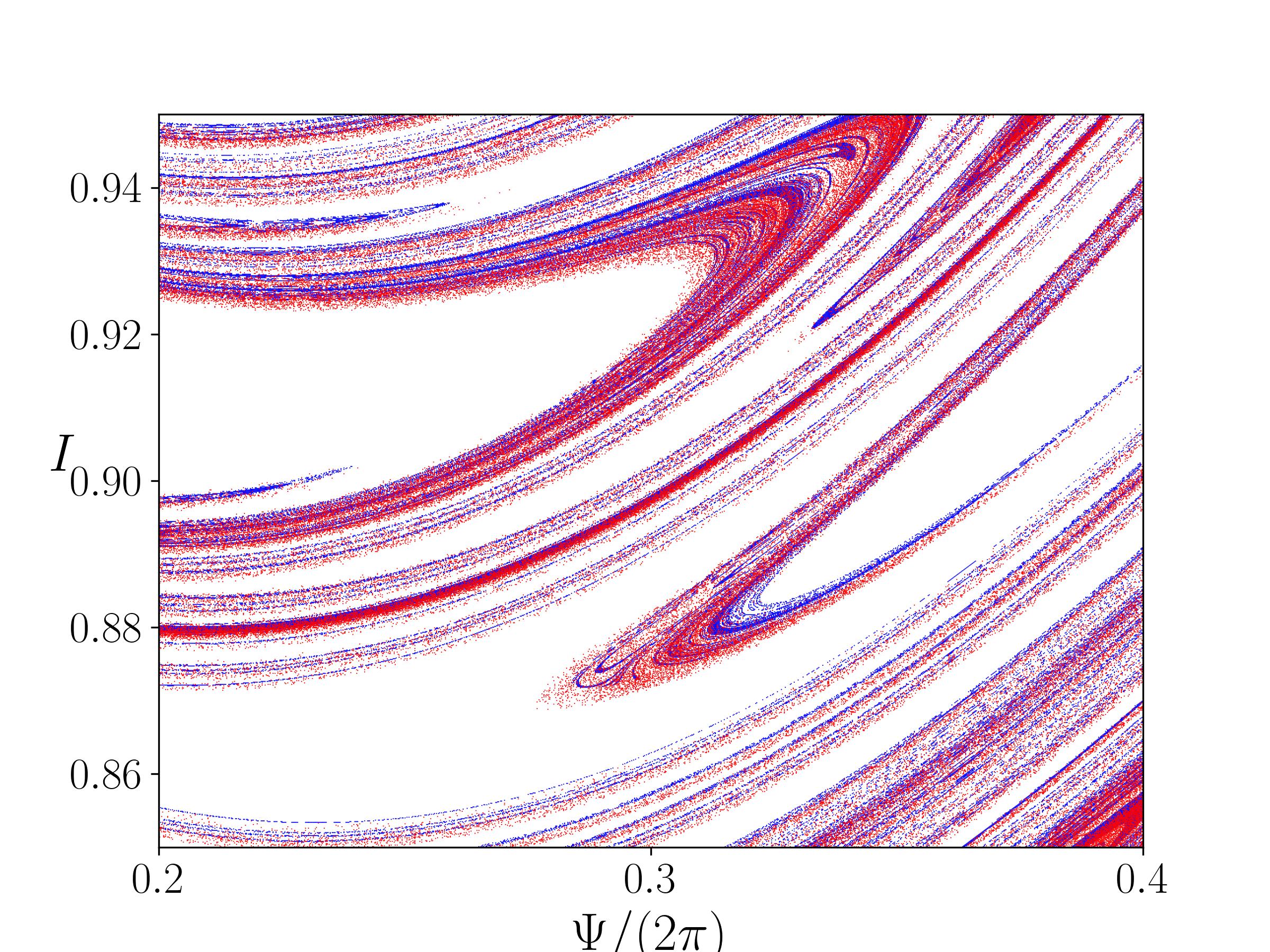}}
    \subfloat(b){\includegraphics[height=2.5in]{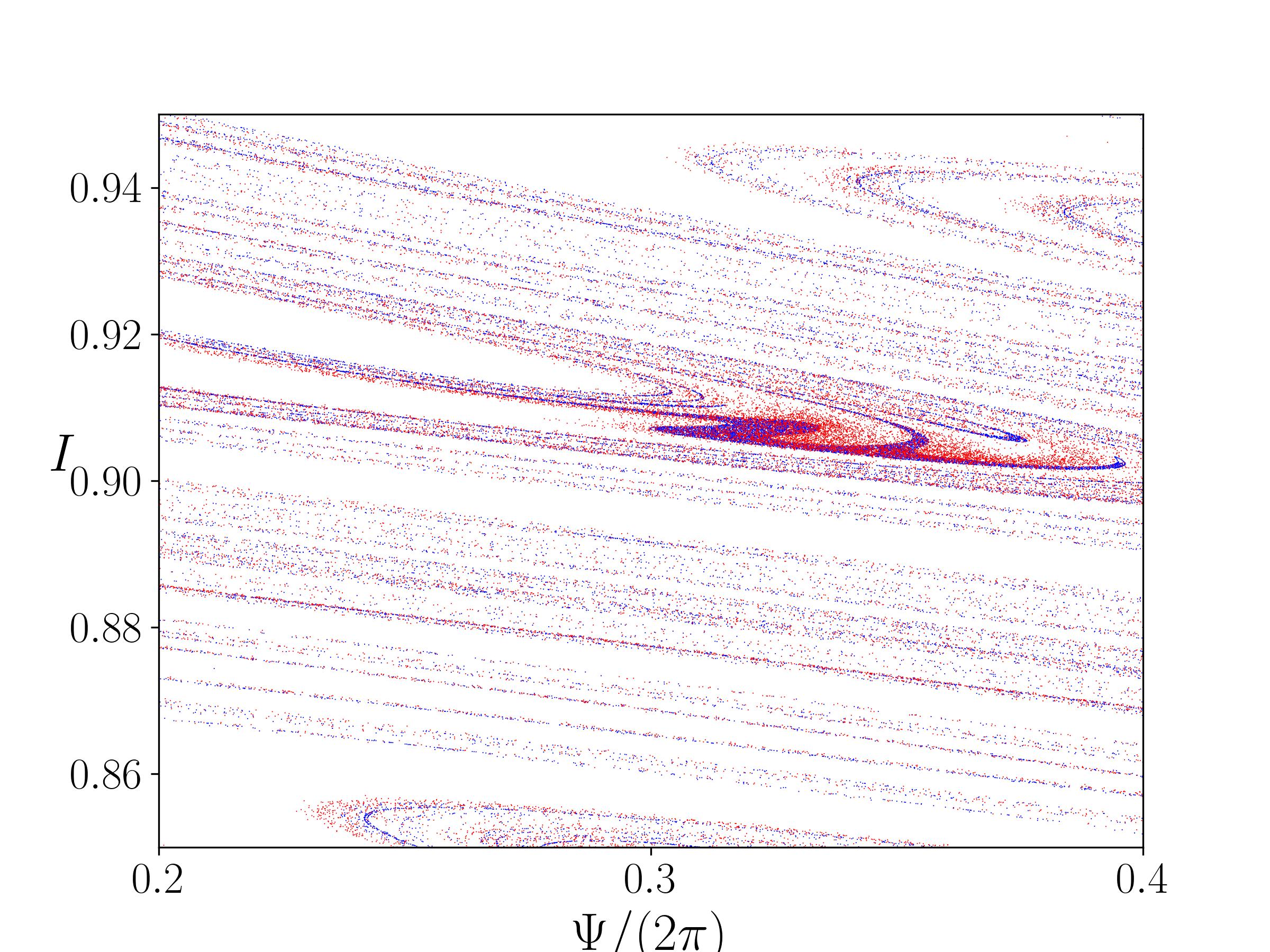}}
    \caption{Numerical approximation of the (a) the stable, and (b) the unstable, manifolds. The blue is for the collisionless map, and in red is for the collision case.}
    \label{fig:colission}
\end{figure}

Random perturbations in the particle motion can lead to the escape of particles, otherwise trapped \cite{seoane,grebogi1}. However, in scattering systems noise can enhance the trapping of trajectories \cite{noise-melhora}. To understand how the escape of particle is affected by collisions, we consider a simple noise model in which we introduced a noise component in our map ${\bf M}({\bf v}_n)$. The collisional effects are included in the map by adding, with a probability $P$, a displacement of size \begin{equation}
\rho = {\bf v}_{n+1} = {\bf M}({\bf v}_n) + {\bf C}^P(\rho),    
\end{equation}
where $C_I^P = \rho\sin(\Gamma)$ and $C_{\Psi}^P = \rho\cos(\Gamma)$, $P$ is the collision probability from a given distribution function and $\Gamma$ is the angle. For each initial condition and at each given instant in time $n$, a new value of $\Gamma$ is randomly generated within the range of $-\pi\leq\Gamma \leq\pi$. With this, the map is non-autonomous as it fully depends of the snapshot (value of $n$). Furthermore it is no longer exactly area preserving. Fig. \ref{fig:colission}  shows the stable (a) and unstable manifolds (b) for the collisionless and collisional map, with $P=1$ and $\rho=10^{-3}$, calculated using the sprinkler method with $1000\times 1000$ initial conditions and $m=10$. In the collision map, the particles do not trace the unstable manifold, rather they disperse about it. The fractal structures can be destroyed in the presence of collisions.

In other words, the main effect of collisions, in terms of particle escape, is the blurring of the stable and unstable invariant manifolds. Instead of Cantor-like sets of filaments, they become a spread of points.



\section{Conclusions}

Chaotic particle transport in magnetized plasmas is a subject of utmost interest in view of its applications in the diffusion of impurities in tokamaks, for example. Charged impurities can be treated as passive tracers, advected by a time-dependent ${\bf E}\times{\bf B}$ flow. From this point of view, particle dynamics can be cast into a Hamiltonian system with one-and-a-half degrees of freedom. As long as we consider the evolution of a limited number of particle impurities, our model would be preferable to a kinetic description, for example. In addition, the Hamiltonian nature of the equations is useful to explain the formation and evolution of a chaotic region near the peripheral region of the tokamak. 

In this work, we investigated the escape of chaotic particle orbits using an area-preserving Poincaré map obtained from a drift Hamiltonian, using realistic profiles and parameters. While many related works focus on statistical properties of particle diffusion, we rather concentrate on the particle dynamics itself, identifying those sets of initial conditions leading to particle escape through exits placed at the tokamak boundary. Those exits can be adapted to include scenarios where divertor plates are suitably placed so as to reduce particle fluxes on sensitive parts of the tokamak inner wall. 

Due to the underlying dynamical structure of the chaotic orbit which leads to particle escape, the escape basins and their boundaries have fractal characteristics, which have been identified and, whenever possible, quantified so as to measure the amount of final-state uncertainty. 

Firstly, we divided the wall into two exits, through which the particles in the chaotic region (in phase space) can escape. The corresponding escape basins and their common boundary are fractal. The structure of the escape basins features an infinite number of fingers which follow the intersection of a basin boundary segment with the stable manifold of an unstable periodic orbit, embedded in the chaotic region. We verified this fact from direct numerical calculation of the invariant manifolds. 

In addition, we quantified the fractality with the box-counting dimension of the escape basin boundary by using the uncertainty exponent method. Our numerical results show a dimension close to the dimension of the phase space itself (equal to two) for a wide interval of the perturbation strength values, and indicating a high degree of fractal behavior. This has important consequences on the predictability of the final state of the system: even if we achieve a great improvement in the uncertainty of the initial condition, this will have nearly no effect on the predictability of the final state of the system. In other words, it is practically impossible to predict by which exit the particle will escape.

Since the values of the box-counting dimension are poorly affected by the intensity of perturbation, we used other quantitative diagnostics of fractal behavior. Accordingly, we calculated the corresponding basin entropy and basin boundary entropy. These quantities may vary between zero, when there is no uncertainty in the final state, and a maximum value of $\log 2$ (in the case of two exits). In the latter case, the basins are so intertwined that, for a randomly chosen particle, the probability of escaping through either exit is the same (equiprobable escape). We found that both entropies increase with the amplitude of the fluctuations, in the same way as the relative area occupied by one of the escape basins. Hence we conclude that this would be a better characterization of fractality than the dimension itself. 

A non-trivial and challenging topological property of fractal basins is the Wada property, for the case of three or more escapes. In our work, we divided the tokamak wall into three exits of the same size, in order to investigate the Wada property, i.e. boundary points having in their neighbourhood points belonging to all three basins. A qualitative way to suggest the existence of Wada property is to show that the unstable manifold stemming from an unstable periodic orbit intersects all basins, what we numerically verified. 

Moreover, a quantitative way to assess the degree to which the Wada property is fulfilled is the grid approach. Using this method of successive refinements, we found that, for a given value of the perturbation strength, $4.0\%$ of the boundary points separate two escape basins, whereas for $96\%$ of the boundary points separate three basins, thus displaying the Wada property, so that the system is partially but almost completely Wada.  

The physical consequences of the Wada property are essentially the same as those deriving from the fractal nature of the escape basin boundaries. The difference, in the former case, is that the concept of fractal boundary, for three or more exits, acquires a more deep and precise meaning, from the mathematical point of view. 

In order to investigate the influence of the collisions on the fractal structures of the particle escape, we added a collisional term to the map, assuming that collisions can be regarded as a noisy component. Within this procedure, we found that collisions make the particles disperse around the invariant manifold rather that trace as the collisionless case.  

The theoretical model we used to describe the ${\bf E}\times{\bf B}$ drift flow, influenced by electrostatic fluctuations, has some evident drawbacks. Firstly we only considered one resonant mode, which is clearly a simplification, given the broadband nature of the measured spectra of electrostatic fluctuations in tokamaks. However, the addition of more resonant modes, while more realistic, would not modify the chaotic region in a way that would affect our main conclusions. Our results are similar to the two drift waves model\cite{amanda-phys-a}, given that the structures studied are consequences of the dynamics underlying chaotic orbits in non-integrable area-preserving systems. Finally, since the theoretical and computational analysis shown in this paper has been applied to ${\bf E}\times{\bf B}$ flows, we speculate that our results may be of interest in other plasma configurations displaying these features, like Hall thrusters \cite{yves2020,yves2021,hall,hall2} and magnetron discharges such as those used for High Power Impulse Magnetron Sputtering (HiPIMS) \cite{magnetrons1}, Penning sources \cite{magnetrons2}, and cusped-field thrusters. 

\section*{Acknowledgments}
The authors would like to thank the anonymous referees who provided useful and detailed comments on a earlier version of the manuscript, and Gabriel Grime for his useful discussions and suggestions. R. L. V. gratefully acknowledges the hospitality extended to him during his stay at the Aix-Marseille University.  The authors thank the financial support from the Brazilian Federal Agencies (CNPq) under Grant Nos. 407299/2018-1, 302665/2017-0, 403120/2021-7, 140713/2020-4, and 301019/2019-3; the São Paulo Research Foundation (FAPESP, Brazil) under Grant Nos. 2018/03211-6 and 2022/04251-7; and support from Coordenação de Aperfeiçoamento de Pessoal de Nível Superior (CAPES) under Grants No. 88887.522886/2020-00, 88881.143103/2017-01 and Comité Français d’Evaluation de la Coopération Universitaire et Scientifique avec le Brésil (COFECUB) under Grant No. 40273QA-Ph908/18.

\end{document}